\PassOptionsToPackage{dvipsnames}{xcolor}

\documentclass[manuscript,screen,nonacm]{acmart}

\AtBeginDocument{%
  \providecommand\BibTeX{{%
    \normalfont B\kern-0.5em{\scshape i\kern-0.25em b}\kern-0.8em\TeX}}}

\setcopyright{none}

\acmJournal{TOCL}

\usepackage{graphicx}
\usepackage{IEEEtrantools}
\usepackage{mathtools}
\usepackage[T1]{fontenc}
\usepackage[utf8]{inputenc}
\usepackage{microtype}
\usepackage{tikz}
\usetikzlibrary{calc,decorations.pathmorphing}
\usepackage{bussproofs}
\usepackage{standalone}

\usepackage{amsmath}
\usepackage{abraces}

\usepackage{hyperref}
\usepackage{algorithm}
\usepackage[noend]{algpseudocode}
\usepackage{comment}

\usepackage{anyfontsize}

\usepackage{thmtools} 

\usepackage{diagbox}
\usepackage{makecell}

\newcommand{\size}{\text{size}}

\newcommand{\Nat}{\mbox{$\mathbb{N}$}}
\renewcommand{\Re}{\mathbb{R}}

\newcommand{\Field}{\mbox{$\mathbb{F}$}}

\newcommand{\viol}{\mathrm{viol}}

\newtheorem{standardlocalcounter}{Dummy}[section]

\theoremstyle{acmplain}

\newtheorem{fact}[standardlocalcounter]{Fact}

\newtheorem{claim}[standardlocalcounter]{Claim}

\newcommand{\set}[1]{\{ #1 \}}

\newcommand{\SET}[1]{\left\{ #1 \right\}}

\newcommand{\setsize}[1]{\lvert#1\rvert}

\newcommand{\union}{\cup}

\newcommand{\bigoh}[1]{\mathrm{O} ( #1 )}

\newcommand{\bigtheta}[1]{\Theta ( #1 )}

\newcommand{\bigomega}[1]{\Omega ( #1 )}

\usepackage{ stmaryrd }

\newcommand{\limpl}{\rightarrow}
\newcommand{\olnot}[1]{\overline{#1}}
\newcommand{\dt}{\mathrm{DT}}
\newcommand{\hpeb}{\mathrm{PebHint}}
\newcommand{\npeb}{\mathrm{Peb'}}
\newcommand{\bpeb}{\mathrm{bpeb}}
\newcommand{\search}[1]{\mathrm{Search}(#1)}
\newcommand{\orfun}{\mathrm{OR}}
\newcommand{\restrict}[2]{{#1}{\upharpoonright_{#2}}}

\newcommand{\ClauseStrut}{\vphantom{$\olnot{x_1}$}}
\renewcommand{\Axiom}[2]{\AxiomC{\textcolor{#1}{\ClauseStrut#2}}}
\newcommand{\AxiomL}[2]{%
  \def\ScoreOverhangLeft{#1}
  \AxiomC{\ClauseStrut}
  \def\ScoreOverhangLeft{4pt}
  \UnaryInfC{\textcolor{Maroon}{\ClauseStrut#2}}
}
\newcommand{\Padding}[1]{\noLine\UnaryInfC{\raisebox{0pt}[\height-.6pt]{\ClauseStrut\phantom{#1}}}}
\renewcommand{\BinaryInf}[1]{\BinaryInfC{\ClauseStrut#1}}

\newcommand{\ceil}[1]{\lceil #1 \rceil}
\newcommand{\floor}[1]{\lfloor #1 \rfloor}

\newcommand{\res}{\textrm{Res}}
\newcommand{\maxres}{\textrm{MaxRes}}
\newcommand{\maxresw}{\textrm{MaxResW}}
\newcommand{\maxresE}{\textrm{MaxResE}}

\newcommand{\tres}{\textrm{TreeRes}}
\newcommand{\SSC}{\textrm{SubCubeSums}}
\newcommand{\DRMaxSAT}{\textrm{DRMaxSAT}}
\newcommand{\PHP}{\textrm{PHP}}

\newcommand{\Int}{\mbox{$\mathbb{Z}$}}

\newcommand{\XOR}{\textrm{XOR}}

\usepackage{relsize}
\newcommand*{\defeq}{\stackrel{\text{{\scriptsize def}}}{=}}

\begin{document}
\title{MaxSAT Resolution and Subcube Sums}
\titlenote{A preliminary version of this article appeared in the proceedings of the 23rd International Conference on Theory and Applications of Satisfiability Testing -- SAT 2020 \cite{FMSV-SAT20}}

\author{Yuval Filmus}
\email{yuvalfi@cs.technion.ac.il}
\orcid{0000-0002-1739-0872}
\affiliation{%
  \institution{Technion -- Israel Institute of Technology}
  \department{Computer Science Department}
  \city{Haifa}
  \country{Israel}}

\author{Meena Mahajan}
\email{meena@imsc.res.in}
\orcid{0000-0002-9116-4398}
\affiliation{%
	\institution{The Institute of Mathematical Sciences  (CI of Homi Bhabha National Institute)}
	\streetaddress{IV Cross Road, CIT Campus, Taramani}
	\city{Chennai}
	\country{India}}

\author{Gaurav Sood}
\email{gauravs@imsc.res.in}
\orcid{0000-0001-6501-6589}
\affiliation{%
	\institution{The Institute of Mathematical Sciences  (CI of Homi Bhabha National Institute)}
	\streetaddress{IV Cross Road, CIT Campus, Taramani}
	\city{Chennai}
	\country{India}}

\author{Marc Vinyals}
\email{marcvinyals@gmail.com}
\orcid{0000-0002-1487-445X}
\affiliation{%
	\institution{Technion -- Israel Institute of Technology}
	\department{Computer Science Department}
	\city{Haifa}
	\country{Israel}}

\renewcommand{\shortauthors}{Y.~Filmus et al.}

\begin{abstract}
We study the MaxSAT Resolution (\maxres) rule in the context of certifying
unsatisfiability. We show that it can be exponentially more powerful
than tree-like resolution, and when augmented with weakening (the
system \maxresw), $p$-simulates tree-like resolution. In devising a
lower bound technique specific to \maxres\ (and not merely inheriting
lower bounds from \res), we define a new %
proof system
called the \SSC\ proof system. This system, which $p$-simulates
\maxresw, can be viewed as a special case of the semialgebraic
Sherali--Adams proof system. In
expressivity, it is the integral restriction of conical juntas studied
in the contexts of communication complexity and extension complexity.  
We show that it is not
simulated by \res. Using a proof technique qualitatively different
from the lower bounds that \maxresw\ inherits from \res, we show that
Tseitin contradictions on expander graphs are hard to refute in
\SSC. We also establish a lower bound technique via lifting: for
formulas requiring large degree in \SSC, their \XOR-ification requires
large size in \SSC.
\end{abstract}




\maketitle

\section{Introduction}
\label{sec:intro}
The most well-studied propositional proof system is
Resolution (\res), \cite{Blake37,Robinson65}.
It is a refutational line-based system that operates on
clauses, successively inferring newer clauses until the empty clause is
derived, indicating that the initial set of clauses is
unsatisfiable.
It has just one satisfiability-preserving rule: if
clauses $A \vee x$ and $B\vee \neg x$ have been inferred, then the
clause $A\vee B$ can be inferred. Sometimes it is convenient,
though not necessary in terms of efficiency,
to also allow a weakening rule:
from clause $A$, a clause $A \vee x$ can be inferred.
While there are several lower bounds known for this system, it is
still very useful in practice and underlies many current SAT solvers.

While deciding satisfiability of a propositional formula is
NP-complete, the MaxSAT question is an optimization question, and
deciding whether its value is as given (i.e.\ deciding, given a
formula and a number $k$, whether $k$ clauses can be
simultaneously satisfied but $k+1$ clauses cannot be satisfied) is potentially harder since
it is hard for both NP and coNP. 
A proof system for MaxSAT was proposed in
\cite{BLM-AI2007,LHG-AI08}. This system, denoted MaxSAT Resolution or more
briefly \maxres, operates on multi-sets of clauses.
At each step, two
clauses from the multi-set are resolved and removed. The resolvent, as
well as certain ``disjoint'' weakenings of the two clauses, are added
to the multiset.
The invariant maintained is that for each assignment
$\rho$, the number of clauses in the multi-set falsified by $\rho$
remains unchanged.  The process stops when the multi-set has a
satisfiable instance along with $k$ copies of the empty clause; $k$ is
exactly the minimum number of clauses of the initial multi-set that
must be falsified by every assignment. \cite{BLM-AI2007}

Since \maxres\ maintains multi-sets of clauses and replaces used
clauses, this suggests a ``read-once''-like constraint \cite{BLM-AI2007}. However, this
is not the case; read-once resolution is not even complete
\cite{IM-CCC95}, whereas \maxres\ is a complete system for certifying
the MaxSAT value (and in particular, for certifying unsatisfiability).
One could use the \maxres\ system to certify unsatisfiability, by
stopping the derivation as soon as one empty clause is produced. Such
a proof of unsatisfiability, by the very definition of the system, can
be $p$-simulated by Resolution. (The \maxres\ proof is itself a proof
with resolution and weakening, and weakening can be eliminated at no
cost.) Thus, lower bounds for Resolution automatically apply to
\maxres\ and to \maxresw\ (the augmenting of \maxres\ with an
appropriate weakening rule) as well. However, since \maxres\ needs to
maintain a stronger invariant than merely satisfiability, it seems
reasonable that for certifying unsatisfiability, \maxres\ is
weaker than Resolution. 
(This would explain why, in practice, MaxSAT solvers do not seem
to use \maxres\ -- possibly with the exception of \cite{NB14Maximum},
but they instead  directly call SAT solvers, which use standard
resolution.)
Proving this would require a lower bound
technique specific to \maxres. 

Associating with each clause the subcube of
assignments that falsify it, each \maxres\ step  manipulates
and rearranges multi-sets of subcubes. This naturally leads us to the
formulation of a static proof system that we call the
\SSC\ proof system. This system, by its very definition, $p$-simulates
\maxresw. Associating with each subcube the minimal conjunction of literals (called terms) that is satisfied by all assignments in the subcube, \SSC\ can be viewed as a special  case of the semi-algebraic
Sherali--Adams proof
system 
(see for instance \cite{FKP19Semialgebraic,ALN14NarrowProofs,Berkholz18Relation,AH19Size}).  Given this position in the ecosystem of simple proof systems,
understanding its capabilities and limitations seems an interesting
question.

\subsection*{Our contributions and techniques}
\label{subsec:contrib}
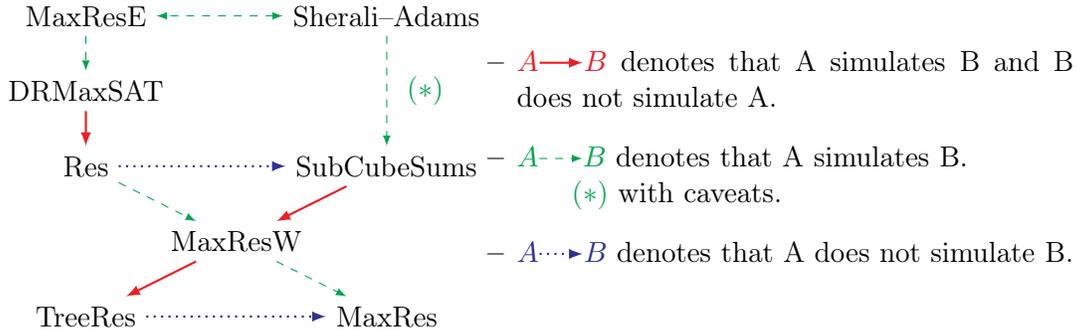
\begin{figure}
\centering
\begin{tikzpicture}
\node (tres) at (0,1) {\tres};
\node (maxres) at (4,1) {\maxres};
\node (maxresw) at (2,2) {\maxresw};
\node (res) at (0,3) {\res};
\node (ssc) at (4,3) {\SSC};
\node (sa) at (4,5) {Sherali--Adams};
\node (drmaxsat) at (0,4) {DRMaxSAT};
\node (maxrese) at (0,5) {MaxResE};

\foreach \i/\j in {ssc/maxresw, maxresw/tres,drmaxsat/res}
\draw[color=Red,-latex,thick] (\i) -- (\j);

\foreach \i/\j in {res/maxresw, maxresw/maxres, maxrese/drmaxsat}
\draw[color=Green,dashed,-latex,thin] (\i) -- (\j);

\draw[color=Green,dashed,-latex,thin] (sa) -- (ssc) node [pos=.5,label=right:$(*)$] {};

\foreach \i/\j in {res/ssc,tres/maxres}
\draw[color=Blue,dotted,-latex,thick] (\i) -- (\j);

\foreach \i/\j in {sa/maxrese}
\draw[color=Green,dashed,latex-latex,thin] (\i) -- (\j);

\node[right] (legend) at (5,3) {
  \parbox{9cm}{
    \begin{itemize}
    \item {\color{Red} $A {\tikz[baseline] \draw[thick,-latex] (0,3pt) to (1.5em,3pt);} B$} denotes that A simulates B
      and B does not simulate A.
    \item {\color{Green} $A {\tikz[baseline] \draw[dashed,thin,-latex] (0,3pt) to (1.5em,3pt);} B$} denotes that A simulates B.\\
    \null\hspace{1.9em}{\color{Green} $(*)$} with caveats.
    \item {\color{Blue} $A {\tikz[baseline] \draw[dotted,thick,-latex] (0,3pt) to (1.5em,3pt);} B$} denotes that A does not simulate B.
    \end{itemize}
} };
\end{tikzpicture}
\caption{Relations among various proof systems}
\label{fig:simulations}
\Description{}
\end{figure}

\begin{enumerate}
\item We observe that for certifying unsatisfiability, the proof
  system \maxresw\ $p$-simulates the tree-like fragment of \res,
  \tres\ (Lemma~\ref{lem:tres-simulation}). This simulation seems to
  make essential use of the weakening rule. On the other hand, we show
  that even \maxres\ without weakening is not simulated by
  \tres\ (Theorem~\ref{thm:tres-separation}). We exhibit a formula,
  which is a variant of the pebbling contradiction~\cite{BW01ShortProofs} on a
  pyramid graph, with short refutations in
  \maxres\ (Lemma~\ref{lem:hpeb-ub}), and
  show that it requires
  exponential size in \tres\ (Lemma~\ref{lem:hpeb-lb}).
\item We initiate a formal study of the newly-defined 
  proof system \SSC. We discuss how it is a natural degree-preserving
  restriction of the  Sherali--Adams proof system
  and touch upon subtleties while defining size.
  We show that the system \SSC\ is not
  simulated by \res, by showing that the Subset Cardinality Formulas, known to be hard for \res, have short \SSC\ refutations (Theorem~\ref{thm:ssc-res}). We also give a direct combinatorial proof that the pigeon-hole principle formulas have short \SSC\ refutations (Theorem~\ref{thm:ssc-php}); this fact is implicit in a recent result from \cite{LR-AAAI20}. 
\item We show that the Tseitin contradiction on an odd-charged
  expander graph is hard for \SSC\ (Theorem~\ref{thm:ssc-Tseitin}) and
  hence also hard for \maxresw. While this already follows from the
  fact that these formulas are hard for
  Sherali--Adams~\cite{AH19Size}, our lower-bound technique is
  qualitatively different; it crucially uses the fact that a stricter
  invariant is maintained in \maxresw\ and \SSC\ refutations.
\item Abstracting the ideas from the lower bound for Tseitin
  contradictions, we devise a lower-bound technique for \SSC\ based on
  lifting (Theorem~\ref{thm:ssc-lifting}).  Namely, we show that if
  every \SSC\ refutation of a formula $F$ must have at least one wide
  clause, then every \SSC\ refutation of the formula $F\circ \oplus$
  must have many cubes. We illustrate how the Tseitin contradiction
  lower bound can be recovered in this way.
\end{enumerate}
The relations among these proof systems are summarized in
Figure~\ref{fig:simulations},
which also includes two proof systems discussed in Related
Work. 

\subsection*{Related work}
\label{subsec:related}
One reason why studying \maxres\ is interesting is that it
displays unexpected power after some preprocessing.
As
described in \cite{IMM-SAT17} (see also \cite{MIM-EPIA17}), the PHP
formulas that are hard for Resolution can be encoded into MaxHornSAT,
and then polynomially many weighted \maxres\ steps
suffice to expose the contradiction. The underlying proof system,
weighted \DRMaxSAT, has been studied further in
\cite{BBIMM-AAAI18}, where it is shown to p-simulate general
Resolution. While  weighted \DRMaxSAT\ gains power from the encoding, the basic
steps are \maxres\ steps. Thus, to understand how unweighted or weighted \DRMaxSAT\ operates,
a better understanding of \maxres\  could be quite useful.
Since \SSC\ can easily refute some formulas hard for Resolution, it
would be interesting to see how DRMaxSAT relates to \SSC.

Some recent papers~\cite{LR-AAAI20,LR-SAT,BL-SAT20} study a generalization
of the weighted version of MaxRes, under names MaxResE and MaxResSV. This 
system allows negative weights in the intermediate steps, as long as all 
the clauses have positive weights at the end. The system is used for certifying
the MaxSAT value in \cite{LR-AAAI20,LR-SAT} and for certifying unsatisfiability in
\cite{BL-SAT20}. This difference allows the system to be used in a slightly 
different way in these papers. Since the satisfiability of a CNF does 
not change if we assign arbitrary positive weights to the axioms, 
\cite{BL-SAT20} allows doing this. On the other hand, this is not allowed 
in \cite{LR-AAAI20,LR-SAT} because this would make the system unsound for 
MaxSAT. With this added power the system in \cite{BL-SAT20} is p-equivalent 
to another recently defined proof system called Circular Resolution \cite{AL19Circular}; hence by the results in \cite{AL19Circular}, it is also p-equivalent to Sherali--Adams. Though most 
results in \cite{LR-AAAI20} are for general MaxSAT, there is one result for 
a special case of MaxSAT where all axioms have infinite weight. Because of
infinite weights, we get a result similar to that
in \cite{BL-SAT20}: the system is p-equivalent to Circular Resolution and 
Sherali--Adams. As can be seen from \cite{BL-SAT20}, the restriction of Circular Resolution where axioms can be used only once is precisely \maxresw; the further restriction of disallowing weakening of axioms is \maxres. 

It is also worth noting that
MaxResW appears in~\cite{LR-SAT} as MaxRes with a split rule, or
ResS. It is shown in \cite{LR-AAAI20,LR-SAT} that for certifying
the MaxSAT value (that is, the optimization version), weakening
provably adds power to \maxres. However, whether weakening adds power
when \maxres\ is used only to certify unsatisfiability remains unclear.

In the setting of communication complexity and of extension complexity
of polytopes, non-negative rank is an important and useful measure. As
discussed in \cite{GLMWZ16Rectangles}, the query-complexity analogue
is {\em conical juntas}; these are non-negative combinations of
subcubes. Our \SSC\ refutations are a restriction of conical juntas to
non-negative {\em integral} combinations. Not surprisingly, our lower
bound for Tseitin contradictions is similar to the conical junta
degree lower bound established in \cite{GJW18ExtensionComplexity}.

Recently, in \cite{FGGR-ITCS22}, one of the open
problems raised in this paper is resolved; a lower bound for
\SSC\ size is shown for a formula that has short refutations in
resolution. Also, in \cite{GHJMPRT-22}, a very close variant of
\maxresw\ called reversible resolution is studied and separated from
resolution. This system has the weakening rule and its reverse; that is, resolution is permitted only when the antecedent clauses  differ in only one variable, which they have in opposing polarities. 

\subsection*{Organisation of the paper}
We define the proof systems \maxres, \maxresw, and \SSC\ in
Section~\ref{sec:defs}.  In Section~\ref{sec:treeres} we relate them
to \tres. In Section~\ref{sec:ssc}, we focus on the \SSC\ proof
system, showing the separation from
\res\ (Section~\ref{sec:ssc-res}), the lower bound for
\SSC\ (Section~\ref{sec:ssc-lb}), and the lifting technique
(Section~\ref{sec:ssc-lifting}).

\section{Defining the Proof Systems}
\label{sec:defs}
A literal is a variable or its negation. A clause is the disjunction of a set of literals (hence, without repetitions). In particular, if $A$ and $B$ are clauses, then $A \vee B$ denotes the clause that is the disjunction of the literals in A and in B without repetitions. A clause is non-tautologous if it has no pair of contradictory literals ($x$ and $\neg x$). We work only with non-tautological clauses throughout. 

For set $X$ of variables, let $\langle X \rangle$ denote the set of
all total assignments to variables in $X$. For a (multi-) set  $F$ of
clauses, $\viol_F\colon\langle X \rangle \rightarrow \{0\}\cup \Nat$ is the
function mapping $\alpha$ to the number of clauses in $F$ (counted
with multiplicity) falsified by $\alpha$. A (sub)cube is the set of
assignments falsifying a clause, or equivalently, the set of
assignments satisfying a conjunction of literals. 
(We refer to clauses and cubes interchangeably, given the natural
bijection between them.) 
The width of a clause is the number of literals in it, and the width of a (multi-) set $F$ of clauses is the maximum width of the clauses it contains. 

The proof system \res\ has the resolution rule inferring $C\vee D$
from $C\vee x$ and $D \vee \overline{x}$, and optionally the weakening
rule inferring $C \vee x$ from $C$ if $\overline{x}\not\in C$. A
  refutation of a CNF formula $F$ is a sequence of clauses $C_1,
  \ldots , C_t$ where each $C_i$ is either in $F$ or is obtained from
  some $j,k <i$ using resolution or weakening, and where $C_t$ is the
  empty clause. The underlying graph of such a refutation has the
  clauses as nodes, and directed edge from $C$ to $D$ if $C$ is used
  in the step deriving $D$. The proof system \tres\ is the fragment of
  \res\ where only refutations in which the underlying graph is a tree
  are permitted. A proof system $P$ simulates ($p$-simulates) another
  proof system $P'$ if proofs in $P$ can be transformed into proofs in
  $P'$ with polynomial blow-up (in time polynomial in the size of the
  proof). See, for instance,  \cite{BIW04Near-optimalSeparation}, for more details.

\subsection{The \maxres\ and \maxresw\ proof systems}
The MaxSAT resolution (\maxres) proof system operates on multi-sets of clauses, and uses the multi-output MaxSAT resolution (\maxres) rule
\cite{BLM-AI2007}, defined as follows:

\[
\begin{array}{ll}
  x \vee a_1 \vee \ldots \vee a_s & (x \vee A) \\
  \overline{x} \vee b_1 \vee \ldots \vee b_t &  (\overline{x} \vee B) \\ \hline
  a_1 \vee \ldots \vee a_s \vee b_1 \vee \ldots \vee b_t &
  \textrm{(the ``standard resolvent'')}\\[2mm]
  \begin{rcases}
  x \vee A \vee \overline{b}_1 \\
  x \vee A \vee b_1 \vee \overline{b}_2 \\
  \vdots  \\
  x \vee A \vee b_1 \vee \ldots \vee b_{t-1} \vee \overline{b}_t
  \end{rcases} &\textrm{(weakenings of $x \vee A$)}
   \\[10mm]
  \begin{rcases}
  \overline{x} \vee B \vee \overline{a}_1 \\
  \overline{x} \vee B \vee a_1 \vee \overline{a}_2 \\ 
  \vdots  \\
  \overline{x} \vee B \vee a_1 \vee \ldots \vee a_{s-1} \vee \overline{a}_s 
\end{rcases} &\textrm{(weakenings of $\overline{x} \vee B$)}
\end{array}
\]
The weakening rule for MaxSAT resolution replaces a clause $A$ by
the two clauses 
$A\vee x$ and $A \vee \overline{x}$. 
While applying either of these rules, the antecedents are removed from
the multi-set and the non-tautologous consequents are added. The
point of the MaxSAT resolution rule is that if $F'$ is obtained
from $F$ by applying these rules, then $\viol_F$ and $\viol_{F'}$ are
the same function.

In the proof system \maxres, a refutation of $F$ is a
sequence $F=F_0, F_1, \ldots , F_s$ where each $F_i$ is a multi-set of
clauses, each $F_i$ is obtained from $F_{i-1}$ by an application of
the MaxSAT resolution rule, and $F_s$ contains the empty clause
$\Box$. In the proof system \maxresw, $F_i$ may also be obtained from
$F_{i-1}$ by using the weakening rule. The \size\ of the proof is the
number of steps, $s$. In \cite{BLM-AI2007,LHG-AI08}, \maxres\ is
shown to be complete for MaxSAT; i.e.~if any assignment must falsify at least $k$ clauses, then at least $k$ copies of the empty clause can be derived using \maxres. Hence \maxres\ is also complete for
unsatisfiability. 
Since the proof system \maxres\ we consider here is 
a refutation system rather than a system for MaxSAT, we can stop
as soon as a single $\Box$ is derived. 

\subsection{The \SSC\ proof system}
The \SSC\ proof system is a static proof system. For an unsatisfiable CNF
formula $F$ (over variable set $X$), a \SSC\ proof is a multi-set $G$ of clauses (or subcubes) over $X$
satisfying $ \viol_F (\alpha) = 1 +
\viol_G (\alpha)$ for all assignments $\alpha \in \langle X \rangle$. The combinatorial size %
of the proof is the number of clauses in $G$ (counting with multiplicity), and the width 
of the proof is the width of $G$. 

Stated in this form, \SSC\ may not be a proof system in the sense of
Cook-Reckhow \cite{CR-JSL79}, since proofs may not be polynomial-time
verifiable.
However, proofs in $\SSC$ can be verified in randomized polynomial time. To see this, we consider an arithmetization of $\SSC$ proofs.

Let $F$ be a CNF formula with $m$ clauses in
variables $x_1, \ldots, x_n$.  Each clause $C_i$, $i\in [m]$, is
translated into a polynomial equation $f_i=0$. A Boolean assignment 
either satisfies clause $C_i$ and  equation $f_i=0$,
or falsifies clause $C_i$ and satisfies equation $f_i=1$. 
(Encoding $e$: $e(x_j)= (1-x_j)$; $e(\neg x_j)= x_j$;
$e(\bigvee_r\ell_r) = \prod_r e(\ell_r)$. So, e.g., clause $x\vee \neg y
\vee z$ translates to the equation $(1-x)y(1-z)=0$. Note that for any non-tautologous clause, each such polynomial $f_i$ is multilinear and has the form  $p_{A,B}\triangleq\prod_{i\in A}x_i \prod_{j\in B}(1-x_j)$ for disjoint $A,B \subseteq [n]$.) %

Given an alleged $\SSC$ proof $G$ of an $F$ that we wish to verify, define the polynomial
\[
p_0(x) = \sum_{A,B\subseteq [n]: A\cap B \neq \emptyset} \alpha_{A,B}
\prod_{i\in A}x_i \prod_{j\in B}(1-x_j)
\]
where the coefficient $\alpha_{A,B}$ is the number of copies in $G$ of
the clause whose encoding is $p_{A.B}$. Define the polynomial $Q(x) =
-\sum_{i\in [m]} f_i(x) + p_0(x) + 1$. That is,
\[
Q(x) = -\left(\sum_{i\in [m]} f_i(x)\right) + \left(\sum_{A,B\subseteq [n]: A\cap B \neq \emptyset} \alpha_{A,B}
\prod_{i\in A}x_i \prod_{j\in B}(1-x_j)\right) + 1
\]
Note that for any Boolean
assignment $\alpha$ to the variables, $Q(\alpha) =
-\viol_F(\alpha)+\viol_G(\alpha)+1$. Thus $G$ is a $\SSC$ proof for
$F$ if and only if $Q(x)$ vanishes on all Boolean assignments.

Now note that $Q(x)$ has two nice  properties with useful consequences for us:
\begin{enumerate}
\item $Q(x)$ is multilinear. 

  Hence, $Q(x)$ vanishes on all Boolean assignments of and only if
  $Q(x)$ vanishes everywhere; i.e.\ $Q(x)=0$ is a polynomial identity. (See for instance \cite[Ex.~2.23 on
    p.~76]{Jukna-BFC-2012})
\item $Q(x)$ can be computed by an algebraic circuit that has
  $O(n(|F|+|G|))$ binary operations, and has variables or the
  constants $-1,+1$ at the leaves. ($O(n)$ operations to encode each
  copy of each clause, and then $O(|F|+|G|)$ operations to add them all
  up.)

  Hence, whether $Q(x)$ is identically $0$ can be tested by a
  randomized algorithm in time polynomial in $n,|F|,|G|$. (Polynomial
  identity testing can be done, using randomization, in time
  polynomial in the size of the circuit representation; see for
  instance \cite{AroraBarak09}.)
\end{enumerate}

\subsection{\SSC\  as a subsystem of the Sherali--Adams proof system}
The arithmetization of \SSC\ proofs discussed above naturally recalls
to mind the semi-algebraic Sherali--Adams proof system over the reals, typically with integer coefficients.  We
recapitulate below the definition of the proof system and observe
that \SSC\ is a subsystem of a specific type.

A Sherali--Adams proof of unsatisfiability of a CNF formula $F$ is a sequence of polynomials $g_i$, $i\in
[m]$; $q_j$, $j\in [n]$; and a polynomial $p_0$ of the form
\[
p_0 =
\sum_{A,B \subseteq [n]: A \cap B =\emptyset}\alpha_{A,B} p_{A,B} =
\sum_{A,B \subseteq [n]: A \cap B =\emptyset}\alpha_{A,B} \prod_{j\in A}x_j \prod_{j\in
  B}(1-x_j) 
\]
where each $\alpha_{A,B} \ge 0$, such that the following polynomial identity holds:
\[ \biggl(\sum_{i\in [m]} g_if_i\biggr) +
\biggl(\sum_{j\in [n]} q_j(x_j^2-x_j) \biggr) +
p_0 +1  = 0
\]
(As before, the polynomials $f_i$ encode the clauses of $F$. The axioms $x_j^2-x_j=0$ for $j\in [n]$, called the Boolean axioms, are used to restrict the set of assignments to Boolean values.)

Note that each $p_{A,B}$, and hence $p_0$, is multilinear. 
The degree or rank of the proof is the maximum degree of any $g_if_i$,
$q_j(x_j^2-x_j)$, and $p_{A,B}$.

The polynomials $f_i$ corresponding to the clauses of $F$, as well as the 
polynomials $p_{A,B}$ in $p_0$, are conjunctions of literals, thus special kinds of  $d$-juntas
(Boolean functions depending on at most $d$ variables). 
So $p_0$ is a non-negative linear
combination of non-negative juntas, that is, 
in the nomenclature of \cite{GLMWZ16Rectangles}, a {\em conical junta}.

Consider the  following restriction of Sherali--Adams:
  \begin{enumerate}
    \item Each $g_i = -1$.
    \item Each $\alpha_{A,B} \in \Int^{\ge 0}$ (non-negative
      integers).%
    \item Each $q_j = 0$.
  \end{enumerate}
Hence, for some non-negative integral $\alpha_{A,B}$, a proof as restricted above is the following polynomial identity:
  \[ -\sum_{i\in [m]} f_i + \Biggl(\sum_{A,B \subseteq [n]: A \cap B =\emptyset }\alpha_{A,B} \prod_{j\in A}x_j\prod_{j\in B}(1-x_j)\Biggr) + 1 = 0
  \]
  This is exactly the form of the arithmetization of \SSC\ proofs discussed in the previous subsection. That is, any \SSC\ proof gives rise to such a restricted Sherali--Adams proof. The converse is also true -- each such restricted
 Sherali--Adams proof corresponds in a natural way to a \SSC\ proof as follows:
each $p_{A,B}$ in  $p_0$ encodes a 
clause (equivalently,  the  subcube of assignments falsifying the clause).  For each disjoint pair
$A,B\subseteq[n]$, the \SSC\ proof has $\alpha_{A,B}$ copies of the corresponding
clause/sub-cube.

It is worth noting that in this equivalence, when we translate a
\SSC\ proof $G$ of a formula $F$ into a restricted Sherali--Adams
proof, the resulting degree is the maximum of the width of $F$ and the width of
$G$. Conversely, when we translate a restricted Sherali--Adams proof
into a \SSC\ proof, the width of the resulting \SSC\ proof is no more than the
original degree.

\subsubsection*{\SSC: The algebraic view with twinned variables}
A Sherali--Adams system may require large number of monomials for some formulas
simply because a clause $C$ with $w$ negated literals gives rise to a
polynomial $f$ with $2^w$ monomials.  The standard approach to handle
this is to use twinned variables, one variable for each literal (i.e. $\overline{x}$ is a new variable), and
include in the set of Boolean axioms the equations
$1-x_i-\olnot{x_i}=0$.  
This makes no difference to the degree of the proof. 
(The encoding $e$ is modified to $e(x_j)= \olnot{x_j}$; $e(\neg x_j)= x_j$;
$e(\bigvee_r\ell_r) = \prod_r e(\ell_r)$. So, e.g., clause $x\vee \neg y
\vee z$ translates to the equation $\olnot{x}y\olnot{z}=0$.) \\
Thus a Sherali--Adams proof is now a sequence of
polynomials $g_i$, $i\in [m]$; $q_j,r_j$, $j\in [n]$; and a polynomial
$p_0$ of the form
\[
p_0 =
\sum_{A,B \subseteq [n]: A \cap B =\emptyset}\alpha_{A,B} \prod_{j\in A}x_j \prod_{j\in
  B}\olnot{x_j}
\]
where each $\alpha_{A,B} \ge 0$, such that
\[ \biggl(\sum_{i\in [m]} g_if_i\biggr) +
\biggl(\sum_{j\in [n]} q_j(x_j^2-x_j) \biggr) +
\biggl(\sum_{j\in [n]} r_j(1-x_j-\olnot{x_j}) \biggr) +
p_0 +1  = 0
\]
We will use this formulation with twinned variables.

The unary size of a Sherali--Adams proof is the sum of (the absolute values of) the coefficients of the polynomials occurring in the proof. We can also define unary reduced size which excludes the Boolean axioms and the polynomials $q_j$ and $r_j$ above. (We can also define binary size, accounting for coefficient bit-sizes when represented in binary, or monomial size, ignoring coefficient sizes altogether and only counting distinct monomials. All these measures have been considered in the literature in different papers and different contexts; see for instance \cite{GHP02ComplexitySemialgebraicSTACS,ALN16NarrowProofs,LN17TightSizeDegreeBoundsSOS,AH18SizeDegree,AH19Size,FKP19Semialgebraic}. For the purposes of this paper, unary and unary reduced size are most relevant.)
The degree or rank of the proof is the maximum degree of any $g_if_i$,
$q_j(x_j^2-x_j)$, $r_jx_j$ and $p_{A,B}$. 

Now, the restriction where each
$g_i=-1$, each $\alpha_{A,B} \in \Int^{\ge 0}$ (non-negative
integers), 
and each $q_j=0$, 
gives
the \SSC\ proof system; an algebraic \SSC\ proof is a polynomial identity of the form 
\[
-\biggl(\sum_{i\in [m]} f_i\biggr) +
\biggl(\sum_{j\in [n]} r_j(1-x_j-\olnot{x_j}) \biggr) +
\biggl(\sum_{A,B \subseteq [n]}\alpha_{A,B} \prod_{j\in A}x_j \prod_{j\in
  B}\olnot{x_j}\biggr)
+1  = 0.
\]

(To be precise, a \SSC\ proof corresponds to an
equivalence class of Sherali--Adams proofs modulo Boolean axioms).

With this algebraic view of \SSC\ in mind, we can define the \emph{algebraic size} of a \SSC\ proof to be the unary size of the smallest
corresponding Sherali--Adams proof (note that this includes the Boolean axioms and $r_j$). We can also define the \emph{algebraic reduced size} of a \SSC\ proof to be unary reduced size of the smallest corresponding Sherali--Adams proof. With these definitions, the following relations are immediate:

For any \SSC\ proof $G$ of a formula $|F|$,
\[\textrm{(combinatorial size of $G$)} + |F| = \textrm{(algebraic reduced size of $G$)}
\le \textrm{(algebraic size of $G$)}.\]
\[\max\{\textrm{width}(G), \textrm{width}(F)\} = \textrm{(algebraic degree of $G$)}.\]

\subsection{Relating various  measures for \SSC\ and \maxresw}
In the combinatorial view of \SSC, the natural complexity measures are
combinatorial size (number of subcubes) and width. In the algebraic view, there are
two measures for size depending on whether or not we count the
monomials from the Boolean axioms (the contributions from
$r_j(1-x_k-\olnot{x_j})$): algebraic size, and algebraic reduced
size.

In the algebraic view, there are also two measures for degree: (1)~the
usual degree of the Sherali-Adams restriction, and (2)~the conical
junta degree, or the degree of the polynomial $p_0$ alone.  As
discussed above, the degree equals the maximum of the initial formula
width and the \SSC\ proof width, while the conical-junta-degree equals
the \SSC\ width.
\[\textrm{width}(G) = \textrm{(conical-junta-degree of $G$)}.\]

It is worth noting that the combinatorial measures can be significantly
smaller than the algebraic measures. If $F$ is the negation of the complete tautology on $n$
variables, then the \SSC\ proof is the empty set, of combinatorial size and width
0. However, the algebraic degree is $n$, and the algebraic size and algebraic reduced size are $2^n$, simply because of the contribution from the
initial formula.

Strictly speaking we do not know if unary Sherali--Adams (or even Sherali--Adams with size measured as the sum of the binary bit-sizes of all coefficients, that is, the usual Sherali--Adams) 
simulates \SSC\ with respect to combinatorial size; hence the caveat in Figure~\ref{fig:simulations}. (The simulation holds with respect to algebraic size, as well as with respect to degree.) 
However, upper bounds on \SSC\ algebraic size imply upper bounds on
Sherali--Adams unary size, while known lower bounds on Sherali--Adams unary reduced size imply lower bounds on \SSC\ algebraic reduced size. Hence for all
practical purposes we can think as if it did.

The following proposition
shows why the proposed restriction of Sherali--Adams to \SSC\ remains complete, and
gives combinatorial and algebraic  size bounds in terms of \maxresw\ refutation size.
\begin{proposition}
\label{prop:maxresw-ssc}
\SSC\ $p$-simulates \maxresw.

For any unsatisfiable formula with $n$ variables and $m$ clauses, a
\maxresw\ refutation of size $s$ can be converted (in polynomial time) to a 
\SSC\ proof of both combinatorial size and
algebraic size $\bigoh{m+ns}$.
\end{proposition}
\begin{proof}
If an unsatisfiable CNF formula $F$ with $m$ clauses and $n \ge 3$ variables has a 
\maxresw\ refutation with $s$ steps, then this derivation produces
$\{\Box\} \cup G$ where the number of clauses in $G$ is at most
$m+(n-2)s-1$.  (A weakening step increases the number of clauses by
1, without creating an empty clause. A \maxres\ step increases it by at most $n-2$, and creates at most one empty clause.) The subcubes
falsifying the clauses in $G$ give a \SSC\ proof.

The simulation still holds if we measure algebraic size. To see that,
observe that we can simulate a weakening step by introducing at most  5 new monomials; deriving clauses $A\vee x$ and $A\vee \neg x$ from $A$ corresponds to rewriting the monomial $m$ encoding $A$ as $mx + m\olnot{x} + m(1-x-\olnot{x})$. More generally, 
given a monomial $m$ and a set of literals $A=a_1,\ldots,a_s$, the polynomial
\begin{align*}
  W(m,A) & \defeq
  ma_1 + m(1-\olnot{a_1}-a_1) \\&+ m\olnot{a_1}a_2 + m\olnot{a_1}(1-\olnot{a_2}-a_2) \\&+ \cdots \\&+ m\olnot{a_1}\cdots \olnot{a_{s-1}}a_s + m\olnot{a_1}\cdots \olnot{a_{s-1}}(1-\olnot{a_s}-a_s)
\\&+  m\olnot{a_1}\cdots \olnot{a_s}
\end{align*}
is identically equal to $m$. It describes the weakening of $m$ by the literals of $A$ using the twinning axioms, and has algebraic size $4s+1 \le 5s$. 
Further, given monomials $m_A=\olnot{x}\cdot e(A)$ and $m_B=x\cdot e(B)$ encoding clauses $x\lor A$ and $\olnot{x} \lor B$, we can simulate the \maxres\ resolution rule by writing
\begin{align*}
  m_A + m_B &= W(m_A,B\setminus A) - m_A\cdot e(B\setminus A) \\
  & + W(m_B,A\setminus B) -m_B \cdot e(A\setminus B) \\
  & + e(A\union B) \\
  & - e(A\union B)\cdot (1-\olnot{x}-x).
\end{align*}
The algebraic size of this expression is $(4|B\setminus A|+1) + (4|A\setminus B|+1) + 6 \le 8n$. 

Hence we can simulate a weakening step with $5$ monomials and a
resolution step with at most $8n$ monomials.
\end{proof}

In Section~\ref{sec:ssc-res} we establish combinatorial size upper bounds in
\SSC\ for certain formulas.
To show that these upper bounds also apply to algebraic size, we
observe that the measures are equivalent in proofs of constant
positive or negative degree. More formally, defining the positive (negative)
degree of a proof as the degree counting only $x_i$ variables
(resp. $\olnot{x_i}$) in $f_i$ and $p_0$, the following holds.

\begin{proposition}
  \label{prop:ssc-extended-size}
  A \SSC\ proof of combinatorial size $s$ and positive (negative) degree $d$ has
  algebraic size $\bigoh{2^d(\setsize{F}+s)}$.
\end{proposition}

\begin{proof}

  We use the following claim.
  \begin{claim}
    Let $p$ be a polynomial with integer coefficients that 
    \begin{enumerate}
      \item is multilinear, on $2n$ variables $\{x_i,\olnot{x_i}\mid
        j\in[n]\}$,
      \item has $\mathit{ \# mon}(p)=s$ monomials (with repetition, i.e\ when written with coefficients $\pm 1$),
      \item has positive (negative) degree $d$, and
      \item %
      vanishes on all Boolean assignments to the variables. 
    \end{enumerate}
    Then there is a polynomial $q$ of the form
    $\sum_{j\in[n]}r_j(1-x_j-\olnot{x_j})$, with \\
    $\sum_{j\in[n]}
    \mathit{ \# mon}(r_j(1-x_j-\olnot{x_j})) \le 3\cdot(2^{d}-1)\cdot
    s$, such that $p + q = 0$ (here we count the monomials with repetition).
    \end{claim}

  To see why the proposition follows from the claim, consider a
  \SSC\ proof of size $s=\setsize{p_0}$ and positive (negative) degree $d$. It has
  the form $\sum_{i\in[m]} f_i = p_0 + 1$ modulo Boolean (twinning) 
  axioms. Applying the claim to the polynomial $p = -\sum_{i\in[m]}
  f_i + p_0 + 1$, which has $|F|+|p_0|+1$ monomials, we obtain a polynomial $q$ such that
  $-\sum_{i\in[m]} f_i + p_0 + 1 + q$ is a a Sherali--Adams
  representative of size at most $\left(1 + 3\cdot (2^{d}-1) \right)\cdot
  (\setsize{F}+\setsize{p_0}+1)$. 
\end{proof}

  \begin{proof}(of Claim)
  We prove the claim for positive degree; the negative degree argument
  is identical. We proceed by induction on $d$.

  Base case: $d=0$. Then $p$ is multilinear on the $n$ variables
  $\{\olnot{x_i}\mid i\in [n]\}$, and vanishes at all $2^n$
  Boolean assignments to its variables. Since the multilinear
  polynomial interpolating Boolean values on the Boolean hypercube is
  unique, and since the zero polynomial is such an interpolating
  polynomial, we already have $p=0$ and can choose $q=0$.

  Inductive Step: For each monomial in $p$ with positive degree $d$,
  pick a positive variable $x$ in the monomial arbitrarily, and
  rewrite the monomial $mx$ as $m - m \olnot{x} - m(1 - \olnot{x}-x)$.
  So $p$ is rewritten as $p'+q''$, where $q''$ collects the parts $m(1
  - \olnot{x}-x)$ introduced above and $p'$ collects the remaining
  monomials.

  Note that the monomials $m$, $m\olnot{x}$ have positive degree
  $d-1$, so $p'$ is a multilinear polynomial with positive degree at
  most $d-1$. Also, it has at most $2s$ monomials.
  Since $p$ and $q''$ vanish on all Boolean assignments,  so does $p'$.  
  The
  inductive claim applied to $p'$ yields $q' = \sum_{j\in[n]} r'_j
  (1-\olnot{x_j}-x_j)$ such that $p' + q' = 0$. Hence for $q=q'-q''$,
  $p+q=0$.
  The polynomial $q$ is of the desired form
  $\sum_{j\in[n]}r_j(1-x_j-\olnot{x_j})$. Counting monomials, $q''$
  contributes at most $3s$ monomials by construction, and the number
  of monomials contributed by $q'$ is bounded by induction, so
  $\sum_{j\in[n]} \mathit{ \# mon}(r_j(1-x_j-\olnot{x_j})) \le
  3s + 3\cdot(2^{d-1}-1)\cdot 2s = 3\cdot(2^{d}-1)\cdot s$.

  \end{proof}

\SSC\ is also implicationally complete in the following sense.
We say that $f \ge g$ if for every truth assignment $x$, $f(x) \ge g(x)$.

\begin{proposition}
  \label{prop:ssc-implicational-completeness}
  If $f$ and $g$ are polynomials with $f \geq g$, then there are
  subcubes $h_j$ and non-negative numbers $c_j$ such that on the
  Boolean hypercube, $f-g = \sum_j c_j h_j$. Further, if $f,g$ are
  integral on the Boolean hypercube, so are the $c_j$.
\end{proposition}
\begin{proof}
A brute-force way to see this is to consider subcubes of degree $n$,
i.e.~a single point (or assignment). For each $\beta\in \{0,1\}^n$, define
$c_\beta = (f-g)(\beta) \in \Re^{\ge 0}$.
\end{proof}

\section{\maxres, \maxresw, and \tres}
\label{sec:treeres}
Since \tres\ allows reuse only of input clauses, while \maxres\ does
not allow any reuse of clauses but produces multiple clauses at each
step, the relative power of these fragments of \res\ is intriguing. In
this section, we show that \maxres\ with the weakening rule, \maxresw,
$p$-simulates \tres, is exponentially separated from it, and 
even \maxres\ (without weakening) is not simulated by \tres.

\begin{figure}
  \centering
  \begin{tikzpicture}[every node/.style={scale=1},xscale=1.4]
\tikzstyle{axiom}=[shape=rectangle, draw, text centered]
\tikzstyle{bag}=[shape=rectangle, rounded corners, draw, text centered]
\tikzset{execute at begin node=\strut}

\coordinate (up) at (0,1) {}; 
\coordinate (down) at (0,-1) {}; 
\coordinate (right) at (1,0) {}; 
\coordinate (left) at (-1,0) {};
\coordinate (final) at (6,0) {}; 

\node[bag] (Box) at (final) {$\Box$};

\node[bag] (f) at ($(Box)+(left)+(up)$) {$f$};
\node[axiom] (notf) at ($(Box)+(right)+(up)$) {$\overline{f}$};

\node[bag] (note-f) at ($(f)+(left)+(up)$) {$\overline{e}\vee f$};

\node[bag] (d) at ($(note-f)+(left)+(up)$) {$d$};
\node[bag] (notd-note-f) at ($(note-f)+(right)+(up)$) {$\overline{d}\vee \overline{e}\vee f$};
\node[bag] (e) at ($(f)+2*(right)+2*(up)$) {$e$};

\node[bag] (notb-d) at ($(d)+(left)+(up)$) {$\overline{b}\vee d$};
\node[bag] (notc-e) at ($(e)+(left)+(up)$) {$\overline{c}\vee e$};

\node[axiom] (a) at ($(notb-d)+(left)+(up)$) {$a$};
\node[axiom] (nota-notb-d) at ($(notb-d)+(right)+(up)$) {$\overline{a}\vee\overline{b}\vee d$};
\node[axiom] (b1) at ($(d)+(right)+2*(up)$) {$b$};

\node[axiom] (b2)  at ($(notc-e)+(left)+(up)$) {$b$};
\node[axiom] (notb-notc-e)  at ($(notc-e)+(right)+(up)$) {$\overline{b}\vee \overline{c}\vee e$};
\node[axiom] (c) at ($(e)+(right)+2*(up)$) {$c$};

\foreach \i/\j in {a/notb-d, nota-notb-d/notb-d, notb-d/d, b1/d,
  b2/notc-e, notb-notc-e/notc-e, notc-e/e, c/e,
  d/note-f, notd-note-f/note-f, note-f/f, e/f,
  f/Box, notf/Box}
\draw[->] (\i) --(\j); 
\end{tikzpicture}
\caption{A tree-like resolution proof}
\label{fig:pyramid-tree-proof}
\Description{}
\end{figure}
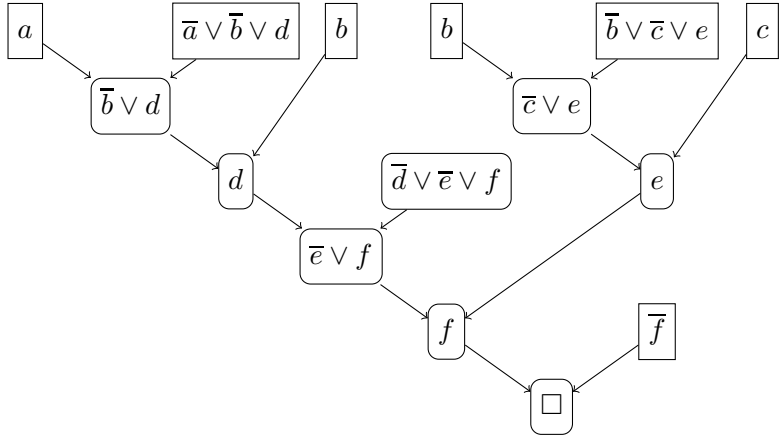

\begin{lemma}
\label{lem:tres-simulation}
For every unsatisfiable CNF $F$, $\size(F\vdash_{\maxresw} \Box) \le 2
\size(F\vdash_{\tres} \Box) $.
\end{lemma}
\begin{proof}
  Let $T$ be a tree-like derivation of $\Box$ from $F$ of size
  $s$. Without loss of generality, we may assume that $T$ is regular \cite{Urquhart95Complexity};
  i.e.~no variable is used as pivot twice on the same path.

  Since a MaxSAT resolution step always adds the standard resolvent,
  each step in a tree-like resolution proof can be performed in
  \maxresw\ as well, provided the antecedents are available.  However,
  a tree-like proof may use an axiom (a clause in $F$) multiple times,
  whereas after it is used once in \maxresw\ it is no longer
  available, although some weakenings are available.
  So we need to work with weaker antecedents.
  We describe below how to obtain sufficient weakenings.

  For each axiom $A \in F$, consider the subtree $T_A$ of $T$ defined
  by retaining only the paths from leaves labeled $A$ to the final
  empty clause. We will produce multiple disjoint weakenings of $A$,
  one for each leaf labelled $A$. Start with $A$ at the final node
  (where $T_A$ has the empty clause) and walk up the tree
  $T_A$ towards the leaves. If we reach a branching node $v$ with
  clause $A'$, and the pivot at $v$ is $x$, weaken $A'$ to $A'\vee x$
  and $A'\vee \overline{x}$. Proceed along the edge contributing $x$
  with $A'\vee x$, and along the other edge with $A'\vee
  \overline{x}$. Since $T$ is regular, no tautologies are created in
  this process, which ends with multiple ``disjoint'' weakenings of $A$. 
  
  After doing this for each axiom, we have as many clauses as leaves
  in $T$. Now we simply perform all the steps in $T$.

  Since each weakening step increases the number of clauses by one,
  and since we finally produce at most $s$ clauses for the leaves, the number
  of weakening steps required is at most $s$.
  \end{proof}
As an illustration, consider the tree-like resolution proof in 
Figure~\ref{fig:pyramid-tree-proof}.
Following the procedure in the proof of the Lemma, the axiom $b$ is
weakened to $b\vee e$ and $b \vee \neg e$, since $e$ is the pivot variable at the branching point where $b$ is used in both sub-derivations.

We now show that even without weakening, \maxres\ has short proofs of
formulas exponentially hard for \tres. We denote the literals $\olnot{x}$ and $x$ by
 $x^0$ and $x^1$ respectively. The formulas that
exhibit the separation are \emph{composed} formulas of the form
$F \circ g$, where $F$ is a CNF formula,
$g\colon\set{0,1}^\ell\to\set{0,1}$ is a Boolean function, there are
$\ell$ new variables $x_1,\ldots,x_\ell$ for each original variable $x$
of $F$, and there is a block of clauses $C \circ g$, a CNF expansion
of the expression $\bigvee_{x^b\in C} (g(x_1,\ldots x_\ell) = b)$,
for each original clause $C\in F$. 
We use the pebbling formulas on single-sink directed acyclic graphs: there is a variable for each node, variables at  sources must be true, the variable at the sink must be false, and at each node $v$, if variables at origins of incoming edges are true, then the variable at $v$ must also be true.

We denote by
$\hpeb(G)$ the standard pebbling formula with additional hints
$u\lor v$ for each pair of siblings $(u,v)$---that is, two
incomparable vertices with a common predecessor---, and we prove the
separation for $\hpeb(G)$ composed with the $\orfun$ function. More
formally, if $G$ is a DAG with a single sink $z$, we define
$\hpeb(G)\circ\orfun$ as follows. For each vertex $v\in G$ there are
variables $v_1$ and $v_2$. The clauses are
\begin{itemize}
\item For each source $v$, the clause $v_1 \lor v_2$.
\item For each internal vertex $w$ with predecessors $u,v$, the
  expression $((u_1\lor u_2)\land(v_1\lor v_2))\limpl(w_1\lor w_2)$,
  expanded into 4 clauses.
\item The clauses $\olnot{z_1}$ and $\olnot{z_2}$ for the sink $z$.
\item For each pair of siblings $(u,v)$, the clause $u_1 \lor u_2 \lor v_1 \lor v_2$. %
\end{itemize}
Note that the first three types of clauses are also present in
standard composed pebbling formulas, while the last type are the hints.

We prove a \maxres{} upper bound for the particular case of pyramid
graphs. Let $P_h$ be a pyramid graph of height $h$ and
$n=\bigtheta{h^2}$ vertices.

\begin{lemma}
\label{lem:hpeb-ub}
The $\hpeb(P_h)\circ\orfun$ formulas have $\bigtheta{n}$ size \maxres{} refutations.
\end{lemma}
\begin{proof}
  We derive the clause $s_1 \lor s_2$ for each vertex $s \in P_n$ in
  layered order, and left-to-right within one layer. If $s$ is a source, then $s_1 \lor s_2$ is readily available as an axiom.
  Otherwise assume that for a vertex $s$ with predecessors $u$ and $v$ and
  siblings $r$ and $t$ -- in this order -- we have clauses
  $u_1 \lor u_2 \lor s_1 \lor s_2$ and $v_1 \lor v_2$, and let us see
  how to derive $s_1 \lor s_2$. (Except at the boundary, we don't have the clause $u_1 \lor u_2$ itself, since it has been used to obtain the sibling $r$ and doesn't exist anymore.) We also make sure that the clause
  $v_1 \lor v_2 \lor t_1 \lor t_2$ becomes available to be used in the
  next step.

  In the following derivation we skip $\lor$ symbols, and we
  colour-code clauses so that \textcolor{OliveGreen}{green} clauses
  are available by induction, axioms are \textcolor{Blue}{blue}, and \textcolor{Maroon}{red} clauses, on the right
  side in steps with multiple consequents, are additional clauses
  that are obtained by the MaxRes rule but not with the usual
  resolution rule.

  \begin{equation*}
    \def\ScoreOverhang{4pt}
    \def\defaultHypSeparation{}
    \Axiom{Blue}{$\olnot{u_1} \olnot{v_1} s_1 s_2$}
    \Axiom{OliveGreen}{$u_1 u_2 s_1 s_2$}
    \BinaryInf{$u_2 \olnot{v_1} s_1 s_2$}
    \Padding{$\olnot{u_1} \olnot{v_1} s_1 s_2$}
    \Padding{$\olnot{u_1} \olnot{v_1} s_1 s_2$}
    \Axiom{Blue}{$\olnot{u_2} \olnot{v_1} s_1 s_2$}
    \insertBetweenHyps{\hskip-30pt}
    \BinaryInf{$\olnot{v_1} s_1 s_2$}
    \AxiomL{60pt}{$u_1 u_2 v_1 s_1 s_2$}
    \Axiom{Blue}{$\olnot{u_1} \olnot{v_2} s_1 s_2$}
    \BinaryInf{$u_2 v_1 \olnot{v_2} s_1 s_2$}
    \Axiom{Blue}{$\olnot{u_2} \olnot{v_2} s_1 s_2$}
    \BinaryInf{$v_1 \olnot{v_2} s_1 s_2$}
    \Axiom{OliveGreen}{$v_1 v_2$}
    \BinaryInf{$v_1 s_1 s_2$}
    \insertBetweenHyps{\hskip-80pt}
    \BinaryInf{$s_1 s_2$}
    \Padding{$s_1 s_2$}
    \Axiom{Maroon}{$v_1 v_2 \olnot{s_1}$}
    \Padding{$v_1 v_2 \olnot{s_1}$}
    \AxiomL{70pt}{$v_1 v_2 s_1 \olnot{s_2}$}
    \Axiom{Blue}{$s_1 s_2 t_1 t_2$}
    \BinaryInf{$v_1 v_2 s_1 t_1 t_2$}
    \insertBetweenHyps{\hskip-78pt}
    \BinaryInf{$v_1 v_2 t_1 t_2$}
    \insertBetweenHyps{\hskip-30pt}
    \noLine\BinaryInfC{}
    \DisplayProof
  \end{equation*}

  The case where some of the siblings are missing is similar: if $r$ is
  missing then we use the axiom $u_1 \lor u_2$ instead of the clause
  $u_1 \lor u_2 \lor s_1 \lor s_2$ that would be available by
  induction, and if $t$ is missing then we skip the steps that use
  $s_1\lor s_2 \lor t_1 \lor t_2$ and lead to deriving
  $v_1 \lor v_2 \lor t_1 \lor t_2$.

  Finally, once we derive the clause $z_1 \lor z_2$ for the sink, we
  resolve it with axiom clauses $\olnot{z_1}$ and $\olnot{z_2}$ to
  obtain a contradiction.

  A constant number of steps suffice for each vertex, for a total
  of $\bigtheta{n}$.
  \end{proof}

We can prove a tree-like lower bound along the lines
of~\cite{BIW04Near-optimalSeparation}, but with some extra care to
respect the hints. As in~\cite{BIW04Near-optimalSeparation} we derive
the hardness of the formula from the \emph{pebble game}, a game where
the single player starts with a DAG and a set of pebbles, the allowed
moves are to place a pebble on a vertex if all its predecessors have
pebbles or to remove a pebble at any time, and the goal is to place a
pebble on the sink using the minimum number of pebbles.
Denote by $\bpeb(P\to w)$ the cost of placing a pebble on a vertex $w$
assuming there are free pebbles on a set of vertices
$P \subseteq V$ -- in other words, the number of pebbles used outside
of $P$ when the starting position has pebbles in $P$. For a DAG $G$
with a single sink $z$, $\bpeb(G)$ denotes
$\bpeb(\emptyset\to z)$. For $U \subseteq V$ and
$v\in V$, the subgraph of $v$ modulo $U$ is the set of
vertices $u$ such that there exists a path from $u$ to $v$ avoiding
$U$.

\begin{lemma}[\cite{Cook74ObservationTimeStorageTradeOff}]
  \label{lem:pyramid-cost}
  $\bpeb(P_h)=h+1$.
\end{lemma}

\begin{lemma}[\cite{BIW04Near-optimalSeparation}]
  \label{lem:peb-intermediate}
  For all $P,v,w$,  we have
  $\bpeb(P\to v) \leq \max(\bpeb(P\to w),\bpeb(P\union\set{w}\to v)+1)$.
\end{lemma}

We deviate slightly from~\cite{BIW04Near-optimalSeparation} and,
instead of directly translating a proof to a pebbling strategy, we go
through query complexity as an intermediate step.
The canonical search problem of a formula $F$ is the relation
$\search{F}$ where inputs are variable assignments
$\alpha\in\set{0,1}^n$ and the valid outputs for $\alpha$ are the
clauses $C\in F$ that $\alpha$ falsifies. Given a relation $f$, we denote by $\dt_1(f)$ the
$1$-query complexity of $f$~\cite{LM19Lifting}, that is the minimum over all decision
trees computing $f$ of the maximum of $1$-answers that the decision
tree receives.\footnote{Essentially the same notion of one-sided query complexity is used in
\cite{PapamakariosR-ICALP22} under the name {\em positive depth}.}

\begin{lemma}
  \label{lem:pebbling}
  For all $G$ we have $\dt_1(\search{\hpeb(G)}) \geq \bpeb(G)-1$.
\end{lemma}

\begin{proof}
  We give an adversarial strategy. Let $R_i$ be the set of variables that are assigned to $1$
  at round $i$. We   initially set $w_0=z$, and  maintain the invariant that
  \begin{enumerate}
  \item there is a distinguished variable $w_i$ and a path $\pi_i$
    from $w_i$ to the sink $z$ such that a queried variable $v$ is $0$ iff
    $v\in\pi_i$; and \label{item:invariant-path}
  \item after each query the number of $1$ answers so far is at least
    $\bpeb(G)-\bpeb(R_i\to w_i)$. \label{item:invariant-cost}
  \end{enumerate}
  
Assume that a variable $v$ is queried. If $v$ is not in the subgraph
  of $w_i$ modulo $R_i$ then we answer $0$ if $v\in \pi_i$ and $1$
  otherwise. Otherwise we consider $p_0=\bpeb(R_i\to v)$ and
  $p_1=\bpeb(R_i\union\set{v}\to w_i)$. By Lemma~\ref{lem:peb-intermediate},
  $\bpeb(R_i\to w_i) \leq \max(p_0,p_1+1)$.
  If $p_0\geq p_1$ then we answer $0$, set $w_{i+1}=v$, and extend
  $\pi_i$ with a path from $w_{i+1}$ to $w_{i}$ that does not contain
  any $1$ variables (which exists by definition of subgraph modulo
  $R_i$). This preserves item~\ref{item:invariant-path} of the
  invariant, and since $p_0 \geq \bpeb(R_i\to w_i)$,
  item~\ref{item:invariant-cost} is also preserved. Otherwise we
  answer $1$ and since $p_1 \geq \bpeb(R_i\to w_i)-1$ the invariant is
  also preserved.

  This strategy does not falsify any  hint clause, because all $0$
  variables lie on a path, or the sink axiom, because the sink is
  assigned $0$ if at all.  Therefore the decision tree ends at a
  vertex $w_t$ that is set to $0$ and all its predecessors are set to
  $1$, hence $\bpeb(R_t\to w_t)=1$. By item~\ref{item:invariant-cost}
  of the invariant the number of $1$ answers is at least $\bpeb(G)-1$.
\end{proof}

To complete the lower bound we use the Pudlák--Impagliazzo
Prover--Delayer game~\cite{PI00LowerBoundDLL} where Prover points to a
variable, Delayer may answer $0$, $1$, or $*$, in which case Delayer
obtains a point in exchange for letting Prover choose the answer, and
the game ends when a clause is falsified.

\begin{lemma}[\cite{PI00LowerBoundDLL}]
  \label{lem:prover-delayer-game}
  If Delayer can win $p$ points, then all \tres{} proofs require size
  at least $2^p$.
\end{lemma}

\begin{lemma}
\label{lem:hpeb-lb}
  $F\circ\orfun$ requires size $\exp(\bigomega{\dt_1(\search{F})})$ in
  tree-like resolution.
\end{lemma}
\begin{proof}
  We use a strategy for the $1$-query game of $\search{F}$ to ensure
  that Delayer gets $\dt_1(F)$ points in the Prover--Delayer game.
  If Prover queries a variable $x_i$ then
  \begin{itemize}
  \item If $x$ is already queried we answer accordingly.
  \item Otherwise we query $x$. If the answer is $0$ we answer $0$,
     otherwise we answer $*$.
  \end{itemize}
  Our strategy ensures that if both $x_1$ and $x_2$ are assigned then $x_1 \lor x_2 = x$. Therefore the game only finishes at a leaf of the decision tree, at which point Delayer earns as many points as $1$s are present in the path leading to the leaf. The lemma follows by Lemma~\ref{lem:prover-delayer-game}.
\end{proof}

The formulas $\hpeb(P_n)\circ \orfun$ are easy to refute in
\maxres\ (Lemma~\ref{lem:hpeb-ub}), but from
Lemmas~\ref{lem:pyramid-cost},\ref{lem:pebbling}, and \ref{lem:hpeb-lb}, they
are exponentially hard for \tres. Hence,
\begin{theorem}
  \label{thm:tres-separation}
\tres\ does not simulate \maxresw\ and \maxres.
\end{theorem}

Note that $\dt_1(f) \leq \dt(f)$ for any relation $f$, therefore
Lemma~\ref{lem:pebbling} also holds for the standard measure of query
complexity. The reason behind using one-sided query complexity is
Lemma~\ref{lem:hpeb-lb}, which is false if we replace $\dt_1$ by
$\dt$. A counterexample is the standard pebbling formula where the
signs of all literals have been flipped, which we denote by
$\npeb(G)$: on the one hand we have that
$\dt(\search{\npeb(G)})=\Omega(n/\log n)$, and on the other hand there
is a tree-like proof of $\npeb(G) \circ \orfun$ of length $\bigoh{n}$.

Alternatively we could use standard query complexity in
Lemma~\ref{lem:hpeb-lb} if we composed our formula with $\oplus$ instead
of $\orfun$, but that would make the upper bound in
Lemma~\ref{lem:hpeb-ub} more intricate.

\section{The \SSC\ Proof System}
\label{sec:ssc}

In this section, we explore the power and limitations of the \SSC\
proof system. On the one hand we show (Theorem~\ref{thm:ssc-res}) that
it has short proofs of the subset cardinality formulas, known to be
hard for resolution but easy for Sherali--Adams.
We also give a direct combinatorial argument to show that the
pigeonhole principle formulas, known to be hard for resolution but
easy in MaxRes with extension, are easy for \SSC. On the other hand we
show a lower bound for \SSC\ for the Tseitin formulas on odd-charged
expander graphs (Theorem~\ref{thm:ssc-Tseitin}). Finally, we establish
a technique for obtaining lower bounds on \SSC\ size: a degree lower
bound in \SSC\ for $F$ translates to a size lower bound in \SSC\ for
$F \circ \oplus$ (Theorem~\ref{thm:ssc-lifting}).

\subsection{\res\ does not simulate \SSC}
\label{sec:ssc-res}
We now show that \res\ does not simulate \SSC. We will give two independent proofs using two different formulas: Subset cardinality formulas and the \PHP\ formulas. The result for \PHP\ formulas is implicit in \cite{LR-AAAI20}, but we provide a new combinatorial proof.

\subsubsection{The Subset Cardinality formulas}
~\\
The first separation is achieved using subset cardinality
formulas~\cite{Spence10Sgen1,VS10ZeroOneDesigns,MN14LongProofs}.
These are defined as follows: we have a bipartite graph
$G(U \union V, E)$, with $\setsize{U}=\setsize{V}=n$. The degree of
$G$ is $4$, except for two vertices that have degree $5$. There is one
variable for each edge. For each left vertex $u\in U$ we have a constraint
$\sum_{e\ni u} x_e \geq \ceil{d(u)/2}$, while for each right vertex $v\in V$ we
have a constraint $\sum_{e\ni v} x_e \leq \floor{d(v)/2}$, both
expressed as a CNF.
In other words, for each vertex $u \in U$ we have the clauses
$\bigvee_{i\in I} x_i$ for
${I\in \binom{E(u)}{\floor{d(u)/2}+1}}$, while for each vertex
$v \in V$ we have the clauses $\bigvee_{i\in I} \olnot{x_i}$ for
${I\in \binom{E(v)}{\floor{d(v)/2}+1}}$.

\begin{theorem}
	\label{thm:ssc-res}
	Subset cardinality formulas have \SSC\ proofs of combinatorial and algebraic size $\bigoh{n}$ but
	require resolution length $\exp(\bigomega{n})$.
\end{theorem}

The lower bound requires $G$ to be an expander, and is proven
in~\cite[Theorem~6]{MN14LongProofs}. The upper bound is the following
lemma.

\begin{lemma}
  \label{lem:ssc-ssc}
  Subset cardinality formulas have \SSC\ proofs of combinatorial and algebraic size $\bigoh{n}$.
\end{lemma}
To obtain the size upper bound, it is convenient to use the algebraic
formulation of \SSC. Our proof below is presented in this
framework. For completeness, we also describe, after this proof, the
direct presentation of the subcubes and a combinatorial argument of
correctness. The combinatorial proof is simply an unravelling of the
algebraic proof, but can be read independently.

\begin{proof}
Our plan is to reconstruct each constraint independently, so that for
each vertex we obtain the original constraints
$\sum_{e\ni u} x_e \geq \ceil{d(u)/2}$ and
$\sum_{e\ni v} \olnot{x_e} \geq \ceil{d(v)/2}$, and then add all of
these constraints together.

Formally, if $F_u$ is the set of polynomials that encode the constraint corresponding to vertex $u$, we want to find suitable subcubes $h_j$ and  write
\begin{equation}
  \label{eq:left-constraints}
  \sum_{f \in F_u} f - \biggl( \ceil{d(u)/2} - \sum_{e\ni u} x_e \biggr) = \sum_j c_{u,j} h_j
\end{equation}
and
\begin{equation}
  \label{eq:right-constraints}
 \sum_{f \in F_v} f - \biggl( \ceil{d(v)/2} - \sum_{e\ni v} \olnot{x_e} \biggr) = \sum_j c_{v,j} h_j
\end{equation}
with $c_{u,j},c_{v,j}\geq 0$ and $\sum_j c_{u,j} = \bigoh{1}$,
so that
\begin{align*}
  \sum_{f \in F}f
  &= \sum_{u \in U} \sum_{f \in F_u} f + \sum_{v \in V} \sum_{f \in F_v} f\\
  &= \sum_{u \in U}\biggl(\ceil{d(u)/2} - \sum_{e\ni u} x_e + \sum_j c_{u,j} h_j\biggr)
   + \sum_{v \in V}\biggl(\ceil{d(v)/2} - \sum_{e\ni v} \olnot{x_e} + \sum_j c_{v,j} h_j\biggr)\\
  &= \sum_{u\in U}\ceil{d(u)/2} + \sum_{v\in V}\ceil{d(v)/2} - \sum_{e \in E} (x_e + \olnot{x_e}) + \sum_j c_jh_j\\
  &= \biggl(1+\sum_{u\in U} 2\biggr) + \biggl(1+\sum_{v\in V} 2\biggr) - \sum_{e\in E} 1 + \sum_j c_jh_j \\
  &= (2n+1) + (2n+1) - (4n+1) + \sum_j c_jh_j  = 1 + \sum_j c_jh_j
\end{align*}
where $c_j = \sum_{v \in U \union V}c_{v,j} \geq 0$.
Hence we can write
  $\sum_{f \in F}f - 1 = \sum_j c_jh_j$ with $\sum_j c_j = \bigoh{n}$.

It remains to show how to derive equations \eqref{eq:left-constraints} and \eqref{eq:right-constraints}. The easiest way is to appeal to the implicational completeness of \SSC, Proposition~\ref{prop:ssc-implicational-completeness}.
We continue deriving equation~\eqref{eq:left-constraints}, assuming for simplicity a vertex of degree $d$ and incident edges $[d]$. Let $\olnot{x_I} = \prod_{i\in I} \olnot{x_i}$, and let
$\SET{\olnot{x_I} : I \in \binom{[d]}{d-k+1}}$ represent a constraint
$\sum_{i\in [d]} x_i \geq k$. Let $f=\sum_{I\in\binom{[d]}{d-k+1}} \olnot{x_I}$ and $g=k-\sum_{i\in[d]} x_i$. For
each point $x \in \set{0,1}^d$ we have that either $x$ satisfies the constraint, in
which case $f(x) \geq 0 \geq g(x)$, or it falsifies it, in which case
we have on the one hand $g(x) = s > 0$, and on the other hand
$f(x) = \binom{d-k+s}{d-k+1}=\frac{(d-k+s) \cdot \cdots \cdot
  s}{(d-k+1) \cdot \cdots \cdot 1} \geq s$.

We proved that $f \geq g$, therefore by
Proposition~\ref{prop:ssc-implicational-completeness} we can write $f-g$ as a
sum of subcubes of size at most $2^d = \bigoh{1}$.

Equation~\eqref{eq:right-constraints} can be derived analogously, completing
the proof for \SSC\ algebraic reduced size, which is the same as combinatorial size.

Since the proof has constant degree, 
Proposition~\ref{prop:ssc-extended-size} implies that combinatorial and algebraic size are at
most a constant factor apart, hence the proof also has algebraic size
$\bigoh{n}$.
\end{proof}

In proving the upper bound in Lemma~\ref{lem:ssc-ssc}, we invoked
implicational completeness from
Proposition~\ref{prop:ssc-implicational-completeness}. However, in our
case the numbers are small enough that we can show how to derive
equation \eqref{eq:left-constraints} explicitly, by solving the
appropriate LP, and without relying on
Proposition~\ref{prop:ssc-implicational-completeness}. As a curiosity,
and in preparation for the combinatorial proof, we display them
next. We have
\begin{align*}
  &\olnot{x_{1,2,3}} + \olnot{x_{1,2,4}} + \olnot{x_{1,3,4}} + \olnot{x_{2,3,4}} - (2-x_1-x_2-x_3-x_4) = \stepcounter{equation}\tag{\theequation}\label{eq:degree4}\\ &2x_1x_2x_3x_4 + x_1x_2x_3\olnot{x_4} + x_1x_2\olnot{x_3}x_4 + x_1\olnot{x_2}x_3x_4 + \olnot{x_1}x_2x_3x_4 + 2 \olnot{x_1x_2x_3x_4}
\end{align*}
and
\begin{align*}
  &\olnot{x_{1,2,3}} + \olnot{x_{1,2,4}} + \olnot{x_{1,2,5}} + \olnot{x_{1,3,4}} + \olnot{x_{1,3,5}} + \olnot{x_{1,4,5}} + \olnot{x_{2,3,4}} + \olnot{x_{2,3,5}} + \olnot{x_{2,4,5}} \\ &+ \olnot{x_{3,4,5}}
  - (3-x_1-x_2-x_3-x_4-x_5) = \stepcounter{equation}\tag{\theequation}\label{eq:degree5} \\
  &2x_1x_2x_3x_4x_5
  + x_1x_2x_3x_4\olnot{x_5} + x_1x_2x_3\olnot{x_4}x_5 + x_1x_2\olnot{x_3}x_4x_5 + x_1\olnot{x_2}x_3x_4x_5\\ &+ \olnot{x_1}x_2x_3x_4x_5
  +2\olnot{x_1x_2x_3x_4}x_5 + 2\olnot{x_1x_2x_3}x_4\olnot{x_5}\\ &+ 2\olnot{x_1x_2}x_3\olnot{x_4x_5} +2\olnot{x_1}x_2\olnot{x_3x_4x_5} +2x_1\olnot{x_2x_3x_4x_5}
  + 7 \olnot{x_1x_2x_3x_4x_5}
\end{align*}

We now give the direct combinatorial proof for the Subset Cardinality Formulas.
The Subset Cardinality Formula SCF says that $G$ has a spanning
subgraph where each $u\in U$ has degree at least 2, the degree-5
vertex in $U$ has degree at least 3, but each $v\in V$ has degree at
most 2.

For $w\in W=U\cup V$, $E_w\subseteq E(G)$ denotes the set of edges incident on $w$.

For a vertex $w$, $f_w$ is the set of clauses enforcing the condition
at vertex $w$, and $F$ is the union of these sets.
A \SSC\ proof should give a clause multiset $H$ such that
\begin{equation}
\label{eqn:SCF}
  \forall \alpha \in \{0,1\}^{|E(G)|}: \viol_F(\alpha) = 1+\viol_H(\alpha).
\end{equation}
In short, $\viol_F = 1+\viol_H$.

We describe such an $H$ whose clauses are also naturally associated
with vertices, so $H$ is the union of clause multisets $h_w$ for each
$w \in W$. The clause sets $f_w$ and $h_w$ are described in Table \ref{table:SCS-SSC-cubes}. 

\begin{table}
  \renewcommand{\arraystretch}{1.25}
  \setlength{\tabcolsep}{6pt}
\begin{tabular}{|l|l|l|l|l|}
\hline
\diagbox{Clause}{Vertex Type}
  & \makecell{$w \in U$ and \\ $\deg(w) = 4$}  & \makecell{$w\in U$ and \\ $\deg(w) = 5$} & \makecell{$w \in V$ and \\ $\deg(w) = 4$}  & \makecell{$w \in V$ and \\ $\deg(w) = 5$}  \\ \hline
  For $A \in {E_w\choose 3}: \bigvee_{e\in A}x_e$  & 1 in $f_w$ & 1 in $f_w$ & & \\ \hline
  For $A \in {E_w\choose 3}: \bigvee_{e\in A}\overline{x_e}$  & & & 1 in $f_w$ & 1 in $f_w$ \\ \hline
$\bigvee_{e\in E_w} x_e$ &  2 in $h_w$ & 7 in $h_w$ & 2 in $h_w$ & 2 in $h_w$ \\ \hline 
$\bigvee_{e\in E_w} \overline{x_e}$ &  2 in $h_w$ & 2 in $h_w$ & 2 in $h_w$ & 7 in $h_w$ \\ \hline
\makecell[l]{For $e\in E_w$:\\ $x_e \vee \bigvee_{f\in E_w\setminus \{e\}} \overline{x_f}$} &
1 in $h_w$ & 1 in $h_w$ &  & 2 in $h_w$ \\ \hline
\makecell[l]{For $e\in E_w$:\\ $\overline{x_e} \vee \bigvee_{f\in E_w\setminus \{e\}}x_f $} &
& 2 in $h_w$ & 1 in $h_w$ & 1 in $h_w$ \\ \hline
\end{tabular}
\caption{The sets $f_w$ and $h_w$: The entries give the multiplicity of the clause in the clause sets depending on the type of vertex $w$.}
\label{table:SCS-SSC-cubes}
\end{table}

Towards proving Equation~\ref{eqn:SCF}, we introduce clause multisets  $f'_w$ and $h'_w$, described in Table \ref{table:SCS-SSC-auxillary-cubes}. (They are not part of the \SSC\ proof.) %
Note that $h'_w$ has only empty clauses, so every assignment falsifies all clauses in all the $h'_w$ put together, totalling $4n+2$.
The $f'_w$ clauses together have two clauses per edge $e=(u,v)$: the unit clause $x_e$ in $f'_u$ and the unit clause $\overline{x_e}$ in $f'_v$. Thus every assignment falsifies exactly $|E|=4n+1$ of the clauses in all the $f'_w$ sets put together. 

The multisets $f'_w$ and $h'_w$  are related to the multisets $f_w$ and $h_w$ by  Equation~\ref{eqn:SCF-vertex} below, which can be verified by inspection
(see Equations~\ref{eq:degree4} and~\ref{eq:degree5} for an example).
\begin{equation}
\label{eqn:SCF-vertex}
 \forall \alpha \in \{0,1\}^{E(G)};  \forall w\in W: \viol_{f_w}(\alpha) + \viol_{f'_w}(\alpha) = \viol_{h_w}(\alpha)  + \viol_{h'_w}(\alpha).
\end{equation}
\begin{table}
	\renewcommand{\arraystretch}{1.25}
	\setlength{\tabcolsep}{6pt}
	\begin{tabular}{|l|l|l|l|l|}
		\hline
		\diagbox{Clause}{Vertex Type}
        & \makecell{$w \in U$ and \\ $\deg(w) = 4$}  & \makecell{$w\in U$ and \\ $\deg(w) = 5$} & \makecell{$w \in V$ and \\ $\deg(w) = 4$}  & \makecell{$w \in V$ and \\ $\deg(w) = 5$}  \\ \hline
		For $e\owns w: \overline{x_e}$ & 1 in $f'_w$ &  1 in $f'_w$ &  & \\ \hline 
		For $e\owns w: x_e$ & & & 1 in $f'_w$ &  1 in $f'_w$  \\ \hline 
		$\Box$ & 2 in $h'_w$ & 3 in $h'_w$ & 2 in $h'_w$ & 3 in $h'_w$ \\ \hline
	\end{tabular}
	\caption{The sets $f'_w$ and $h'_w$: The entries give the multiplicity of the clause in the clause sets depending on the type of vertex $w$.}
	\label{table:SCS-SSC-auxillary-cubes}
\end{table}
Hence
\begin{align*}
\viol_F = \sum_{w\in W} \viol_{f_w}
& = \sum_{w\in W} \left( \viol_{h_w} + \viol_{h'_w} - \viol_{f'_w} \right) \\
& = \left( \sum_{w\in W} \viol_{h_w} \right) +
\left( \sum_{w\in W} \viol_{h'_w} \right)
- \left( \sum_{w\in W} \viol_{f'_w} \right) \\
&= \viol_H + (2|U|+1) + (2|V|+1) -
\sum_{e\in E(G)} \left( \viol_{x_e}+ \viol_{\overline{x_e}}\right) \\
&= \viol_H + (4n+2) - (4n+1) = \viol_H + 1
\end{align*}

\subsubsection{The Pigeonhole Principle formulas}

Recall the definition of the Pigeonhole Principle (\PHP) formulas: 
\begin{definition}[$\PHP_m$]
  \label{def:php}
  The clauses of $\PHP_m$ are defined as follows:
  \begin{itemize}
   \item Pigeon axioms -- For each $i\in [m+1]$, $P_i$ is the clause
     $\bigvee_{j=1}^{m} x_{i,j}$
   \item Hole axioms -- For each $j\in [m]$, $H_j$ is the collection of clauses 
     $H_{i,i',j}: \neg x_{i,j} \vee \neg x_{i',j}$ for $1 \le i < i' \le m+1$.
  \end{itemize}
\end{definition}
These formulas are known to be hard for Resolution (\cite{Haken-TCS85}).

In \cite{LR-AAAI20} the authors show that these formulas are easy to
refute in \maxresE, an extended version of \maxres.  This extended
version allows intermediate clauses with negative weights, and,
interpreting $\viol$ as the sum of the weights of the falsified
clauses, rather than merely the number of falsified clauses, all rules
preserve $\viol$. The system allows introducing certain clauses ``out
of nowhere'' preserving this invariant; in particular, it allows the
introduction of triples of weighted clauses of the form $(\Box, -1),
(x,1), (\neg x, 1)$. Consider the following set of clauses, called the
``residual'' of \PHP\ and denoted $\PHP^\delta$:

\begin{definition}[$\PHP^\delta$ from Theorem 5 of \cite{LR-AAAI20}]
  The clause set $\PHP^\delta$ is the set
  \[\bigcup_{i\in[m+1]}P_i^\delta \cup \bigcup_{j\in[m]}H_j^\delta\] where $P_i^\delta$ and $H_j^\delta$ are defined as follows:
  \begin{itemize}
    \item The clause set $P_i^\delta$ encodes
      that pigeon $i$ goes into at most one hole. It is the set
      \[P_i^\delta = \left\{ \neg x_{i,j} \vee \biggl(\bigvee_{j < \ell < k}x_{i,\ell}\biggr) \vee \neg x_{i,k} \,\middle|\, 1 \le j < k \le m\right\}.\]
    \item The clause set $H_j^\delta$ says that
      hole $j$ has at least one and at most two pigeons. It is defined
      as $H1_j^\delta \cup H2_j^\delta$, where
      \begin{itemize}
    \item $H1_j^\delta$ has a single clause encoding that hole $j$ is not empty. 
      \[H1_j^\delta =  \left\{ \bigvee_{i=1}^{m+1} x_{i,j}\right\}.\]
      
    \item $H2_j^\delta$ is a set of clauses encoding that no hole has
      more than two pigeons. It is the set
      \[H2_j^\delta = \left\{\neg
      x_{i,j} \vee
      \biggl(\bigvee_{i<\ell<k}x_{\ell,j}\biggr) \vee \neg x_{k,j} \vee \neg x_{i',j}
      \,\middle|\, 1 \le i < k < i' \le m+1\right\}.\]
      \end{itemize}

  \end{itemize}
\end{definition}

\begin{theorem}[implicit in \cite{LR-AAAI20} Theorem 5]
\label{thm:ssc-php}
  $\viol_{\PHP^\delta}=\viol_{\PHP}-1$.
\end{theorem}

  In the proof of Theorem 5 in \cite{LR-AAAI20}, a \maxresE\ derivation
  transforming $\PHP$ to $\PHP^\delta \cup \{\Box\}$ is
  described. Each step in the derivation preserves the weighted sum of
  violations. (At intermediate stages, some clauses have negative
  weight, hence weighted sum.)

  More precisely, the three weighted clauses $(\Box, -1), (x,1), (\neg
  x, 1)$ have weighted $\viol=0$: Every assignment falsifies one of
  the unit clauses with weight $+1$ and falsifies the empty clause with
  weight $-1$, so the total weight of falsified clauses is 0.  The
  derivation in \cite{LR-AAAI20} adds $m$ such triples. It uses the
  weighted-$\viol$-preserving rules of \maxresE\ to transform $\PHP_m
  \cup \{(\Box,-m) \} \cup \{ x_{1,j}, \neg x_{1,j} \mid j\in [m]\}$
  to $\PHP^\delta \cup \{\Box\}$. Here all clauses of $\PHP_m$ initially have
  weight 1, and all clauses of $\PHP^\delta$ finally have weight 1. Thus the
  proof establishes the following statement:

\begin{corollary}
  \label{cor:ssc-php}
  $\PHP_m$ has a \SSC\ refutation of combinatorial size polynomial in $m$. 
\end{corollary}
\begin{proof}
The cubes falsifying the $O(m^4)$ clauses of $\PHP^\delta$ are the $\SSC$
refutation of $\PHP_m$.
\end{proof}

In \cite{LR-AAAI20} the authors say (just before Theorem 5 and in the
footnote) that it is not obvious that the refutation is complete
though we know this because $\PHP_m$ is minimally unsat.  Actually the
fact that $\PHP^\delta$ is satisfiable is obvious: the assignment that
sets $x_{i,i}=1$ for $i\in[m]$ and all other variables to 0 satisfies
$\PHP^\delta$. (Any matching of size $m$ satisfies $\PHP^\delta$.)
Thus, since $\PHP$ is minimally unsatisfiable, the MaxSAT value of
$\PHP$ and $\{\Box\} \cup \PHP^\delta$ is the same.  However, it is not
obvious why $\viol_{\PHP^\delta}=\viol_{\PHP}-1$.  We show how to prove this
directly without using the \maxresE\ derivation route.  For every
assignment $A$ to the variables of $\PHP$, we show below that
$\viol_\PHP(A) = \viol_{\PHP^\delta}(A)$.

\begin{enumerate}
  \item Let $A\in\{0,1\}^{(m+1)\times m}$ be an assignment to the
    variables of $\PHP_m$.
  \item Denote the column-sums by $c_j = \sum_{i\in[m+1]} A_{i,j}$ for
    $j\in[m]$.
  \item Denote the row-sums by $r_i = \sum_{j\in[m]} A_{i,j}$ for
    $i\in[m+1]$.
  \item Denote the total sum by $M$; $M=\sum_i r_i = \sum_j c_j$. 
\end{enumerate}
It is straightforward to see that 
\[\viol_\PHP(A) = \#\{i\in[m+1]: r_i=0\} + \sum_{j\in[m]} {c_j\choose 2}.\]

To describe $\viol_{\PHP^\delta}(A)$, consider the three sets of
clauses separately.
\begin{enumerate}
\item For pigeon $i$, if $r_i=0$ or $r_i=1$, then there are no
  violations in $P_i^\delta$ since each clause has two negated literals.

  If $r_i\ge 2$, %
  let the positions of the 1s in the $i$th row be $j_1, j_2, \ldots ,
  j_{r_i}$ in increasing order. Then the only clauses falsified are of
  the form
  \[\neg x_{i,j_p} \vee \left(\bigvee_{\ell =
    j_p+1}^{j_{p+1}-1}x_{i,\ell}\right) \vee \neg x_{i,j_{p+1}}\]
  for $p \in [r_i - 1]$, and all these clauses are falsified. So
  $\viol_{P_i^\delta}(A) = r_i-1$.
  
\item The clause in $H1_j^\delta$ is falsified iff $c_j=0$.  

  \item For hole $j$, if $c_j \le 2$, then there
    are no violations in $H2_j^\delta$ since each clause has three
    negated literals.

    If $c_j \ge 3$, then suppose the 1s are in positions $i_1, i_2,
    \ldots , i_{c_j}$ in increasing order.  Then the clauses violated
    are exactly those of the form
      \[\neg x_{i_q,j} \vee
      \left(\bigvee_{i= i_q+1}^{i_{q+1}-1} x_{i,j}\right) \vee \neg
      x_{i_{q+1},j} \vee \neg x_{i_{q+1+k},j}\] for $q,k \ge 1$ and  $q+1+k \le c_j$.  So the number of violations is
      $(c_j-2) + (c_j-3) + \ldots + 1 = {c_j-1 \choose 2}$.  
\end{enumerate}

Putting this together, we have 
\[
\viol_{\PHP^\delta}(A) = \sum_{i\in[m+1]: r_i \ge 2} (r_i-1)
+ \#\{j\in[m]: c_j=0\} + 
\sum_{j\in[m]: c_j \ge 3} {c_j-1\choose 2}.\]
Consider the following manipulations:
\begin{eqnarray*}
  \sum_{i\in[m+1]: r_i \ge 2} (r_i-1) &=&
  \sum_{i\in[m+1]} (r_i-1)
  -   \sum_{i\in[m+1]: r_i =0} (r_i-1)
\\
  &=& 
  \left( \sum_{i\in[m+1]} r_i
  - \sum_{i\in[m+1]} 1 \right)  - \Bigg( (-1)\times \textrm{number of 0-rows} \Bigg) \\
  &=& M - (m+1) + \textrm{number of 0-rows}
\end{eqnarray*}
\begin{eqnarray*}
  \sum_{j\in[m]: c_j \ge 3} {c_j-1\choose 2} &=&
    \sum_{j\in[m]: c_j \ge 1} {c_j-1\choose 2} =
    \sum_{j\in[m]: c_j\ge 1} \left[ {c_j\choose 2} - (c_j-1)\right]\\
&=&  \sum_{j\in[m]: c_j\ge 1} {c_j\choose 2} - \sum_{j\in[m]: c_j \ge 1}(c_j-1)\\
    &=& \sum_{j\in[m]} {c_j\choose 2} - \sum_{j\in[m]} c_j + \sum_{j\in[m]: c_j \ge 1} 1 \\
    &=& \sum_{j\in[m]} {c_j\choose 2} - M + (m- \textrm{number of 0-columns}) \\
\end{eqnarray*}
Putting this together, we obtain 
\begin{eqnarray*}
  \viol_{\PHP^\delta} &=& \sum_{i\in[m+1]: r_i \ge 2} (r_i-1)
+ \#\{j\in[m]: c_j=0\} + 
\sum_{j\in[m]: c_j \ge 3} {c_j-1\choose 2} \\
&=& M - (m+1) + \textrm{number of 0-rows}  \\
&& + \textrm{number of 0-columns} \\
&& +  \sum_{j\in[m]} {c_j\choose 2} - M + (m- \textrm{number of 0-columns}) \\
&=& \textrm{number of 0-rows} +  \sum_{j\in[m]} {c_j\choose 2} - 1 \\
&=& \viol_\PHP - 1
\end{eqnarray*}
as claimed.

In particular, we have the identity:
\begin{proposition}
  \label{prop:identity}
  For any $A \in \{0,1\}^{(m+1)\times m}$,
with row sums $r_i = \sum_j A_{i,j}$ and column sums $c_j = \sum_i A_{i,j}$, 
\begin{eqnarray*} 
&& \#\{i\in[m+1]: r_i=0\} + \sum_{j\in[m]} {c_j\choose 2} \\
& = & 1 + \#\{j\in[m]: c_j=0\} + \sum_{i\in[m+1]: r_i \ge 2} (r_i-1)+
\sum_{j\in[m]: c_j \ge 3} {c_j-1\choose 2}
\end{eqnarray*}
\end{proposition}

We can improve Corollary~\ref{cor:ssc-php}
to a stronger claim about algebraic size.
\begin{corollary}
  \label{cor:ssc-php-esize}
  $\PHP_m$ has a refutation in \SSC\ with  algebraic size polynomial in $m$. 
\end{corollary}
\begin{proof}
Viewing the \SSC\ proof in Corollary~\ref{cor:ssc-php} from the
algebraic viewpoint, the degree of the proof is linear. However, the
negative degree is $3$. So we can still use
Proposition~\ref{prop:ssc-extended-size} to conclude that there is a
refutation with algebraic size $\bigoh{m^4}$.  \end{proof}

\subsection{A lower bound for \SSC}
\label{sec:ssc-lb}
Fix any graph $G$ with $n$ nodes and $m$ edges, and let $I$ be the
node-edge incidence matrix. Assign a variable $x_e$ for each edge $e$.
Let $b$ be a vector in $\{0,1\}^n$ with $\sum_i b_i \equiv 1 \bmod
2$. The Tseitin contradiction asserts that the system $IX=b$ has a 
solution over $\Field_2$. The CNF formulation has, for each vertex $u$ in $G$, with
degree $d_u$, a set $S_u$ of $2^{d_u-1}$ clauses expressing that the
parity of the set of variables $\{x_e \mid \textrm{$e$ is incident on
	$u$} \}$ equals $b_u$.

For these formulas, \res\ refutations  require exponential size \cite{Urquhart87HardExamples},
and hence  \maxresw{} refutations also require exponential size. We now show that  \SSC\ refutations also require  exponential combinatorial size (and hence also algebraic size). By Theorem~\ref{thm:ssc-res}, this lower bound cannot be
inferred from hardness for \res.

We will use these standard facts:
\begin{fact}\label{fact:Tseitin}
	For connected graph $G$, over $\Field_2$,
	\begin{enumerate}
		\item if $\sum_ib_i \equiv 1 \bmod 2$, then the equations $IX=b$
		have no solution.
		\item If $\sum_ib_i \equiv 0 \bmod 2$, then
		$IX=b$ has exactly $2^{m-n+1}$ solutions.
		\item Furthermore, for any assignment $a$, and any vertex $u$, $a$
		falsifies at most one clause in $S_u$.
	\end{enumerate}
	
\end{fact}

A graph is a $c$-expander if for all $V' \subseteq V$ with $|V'| \le
|V|/2$, $|\delta(V')| \ge c |V'|$, where $\delta(V') = \{ (u,v) \in E \mid u \in V', v \in V \setminus V' \}$. 

\begin{theorem}
	\label{thm:ssc-Tseitin}
	Let $G$ be a $d$-regular $c$-expander on $n$ vertices where $n$ is
	odd, and $c,d$ be constants with $c > 10$. Let $b$ be the all-1s vector.
	All \SSC\ refutations of the Tseitin contradiction corresponding to $G,b$
	require combinatorial size exponential in $n$.
\end{theorem}

We prove this using the combinatorial view of \SSC.  At a high level,
the proof proceeds as follows. The Tseitin contradiction $F$ has
$m=dn/2$ variables and $n2^{d-1}$ clauses. The assignments can be
partitioned into disjoint sets $X_i$, where $X_i$ consists of
assignments falsifying exactly $i$ clauses of $F$.  By
Fact~\ref{fact:Tseitin}, $X_i$ is empty for even $i$. We focus on $X_1$, $X_3$, and $X_5$ for the lower bound. 

Let $\mathcal{C}$ be a \SSC\ refutation of $F$, that is,
$\viol_{\mathcal{C}} = \viol_F-1 = g$.  Define a matrix $M$ with rows
indexed by assignments to variables and columns indexed by
clauses/cubes of $\mathcal{C}$, and entries as follows.
\[ M(a,C) = \left\{\begin{array}{ll}
1 &\textrm{~if $a$ falsifies $C$}\\
0 &\textrm{~otherwise}
\end{array}\right .
\]
For each $a\in X_i$, row $a$ of $M$ has exactly $(i-1)$ 1s. Thus the
submatrix $X_3 \times \mathcal{C}$ has $2|X_3|$ 1s, and the submatrix
$X_5 \times \mathcal{C}$ has $4|X_5|$ 1s. We say that a clause is
heavy if it contributes many more 1s in the $X_5$ rows than in the
$X_3$ rows; otherwise it is light.

The proof idea is to show that a significant fraction of the 1s in
$X_3 \times \mathcal{C}$ come from light clauses
(Lemma~\ref{lem:many-light} below), and that a light clause can
contribute only an exponentially small fraction of the 1s in $X_3
\times \mathcal{C}$ (Lemma~\ref{lem:light-contribution-low} below). It
then follows that $\mathcal{C}$ must have exponentially many light
clauses.

\newcommand{\reld}{\text{rel-density}}

For a clause $C \in \mathcal{C}$, let $N_i(C)$ denote the number of 1s
it contributes to $M$ in the rows corresponding to $X_i$. That is
viewing $C$ as the cube of its falsifying assignments, $N_i(C) =
|C\cap X_i|$.  Define the relative density of a clause $C$, denoted
$\reld(C)$, to be the ratio $N_5(C) / N_3(C)$. Say that a clause is
{\em light} if $\reld(C) \le n^2/9$. That is,
for a light $C$,
\[\reld(C) \triangleq
\frac{\textrm{number of 1s in $X_5 \times \{C\}$}}{\textrm{number of 1s in $X_3 \times \{C\}$}}
\leq \frac{n^2}{9}.
\]
In particular, if $C$ is light, $|C \cap X_3|$ is not zero; hence
there is at least one assignment $a\in X_3$ that falsifies $C$. This
fact will be significant.

\begin{lemma}\label{lem:many-light}
	\[
	\frac{\textrm{number of 1s in $X_3 \times \mathcal{C}$ contributed by light clauses}}{\textrm{number of 1s in $X_3 \times \mathcal{C}$}} \ge \frac{1}{10}\]
\end{lemma}

\begin{lemma}\label{lem:light-contribution-low}
	For a light clause $C\in \mathcal{C}$,
	\[ N_3(C) \triangleq |C\cap X_3| \le \frac{3|X_3|}{2^{n(0.1c-1)}} \]
\end{lemma}

Before proving these lemmas, we show why they imply the theorem.
\begin{proof}(of Theorem~\ref{thm:ssc-Tseitin}, assuming Lemmas~\ref{lem:many-light},\ref{lem:light-contribution-low})
	
	\begin{IEEEeqnarray*}{rClr}
	{2|X_3|} &=& \textrm{(number of 1s in $X_3 \times \mathcal{C}$)} & \\
	&\le& 10 \times \textrm{(number of 1s in $X_3 \times \mathcal{C}$ contributed by light clauses)} & \textrm{(by Lemma~\ref{lem:many-light})} \\
	&\le& \IEEEeqnarraymulticol{2}{l}{
	10 \times \textrm{(number of light clauses)} \times 
	{\textrm{(max number of 1s  contributed by a light clause)}}
	} \\
	&\le&  10 \times |\mathcal{C}| \times \frac{3|X_3|}{2^{n(0.1c-1)}} & \textrm{(by Lemma~\ref{lem:light-contribution-low})} \\
	\textrm{Hence~} |\mathcal{C}| &\ge& \frac{2^{n(0.1c-1)}}{15} = 2^{\Omega(n)}.
	\end{IEEEeqnarray*}
\end{proof}

Here is a simple proposition that will be used in proving both Lemmas.
\begin{proposition}\label{prop:Xi-counts}
	For each  odd $i$, $|X_i| = {n \choose i} 2^{m-n+1}$. 
\end{proposition}
\begin{proof}
	An assignment in $X_i$ lies in $i$ cubes of $f$. Each cube corresponds
	to a distinct vertex because the $2^{d-1}$ cubes corresponding to any
	single vertex are disjoint. Once the $i$ vertices are fixed and $b$
	flipped in those coordinates to get $b'$, there are $2^{m-n+1}$ 0-1
	solutions to $Ix=b'$ (Fact~\ref{fact:Tseitin}(2)).
\end{proof}

Now we prove that many 1s in $X_3\times \mathcal{C}$ are contributed
by light clauses.
\begin{proof}(of Lemma~\ref{lem:many-light})
	Consider the following probability distribution $\mu$ on $\mathcal{C}$:
	\[\mu(C) \triangleq
	\frac{|C \cap X_3|}{\textrm{number of 1s in $X_3 \times \mathcal{C}$}}
	= \frac{|C\cap  X_3|}{2|X_3|}.
	\]
	This distribution is useful because it can be used to neatly express
	the quantity we want to bound from below, as follows:
	\begin{IEEEeqnarray*}{rCl}
		\IEEEeqnarraymulticol{3}{l}{
			\frac{\textrm{number of 1s in $X_3 \times \mathcal{C}$ contributed by light clauses}}{\textrm{number of 1s in $X_3 \times \mathcal{C}$}} 
		}\\ \qquad\qquad\qquad\qquad\qquad\quad\quad\quad
		&=& \frac{\sum_{C\in\mathcal{C}; C\textrm{~{\small light}}}|C\cap X_3|}{2|X_3|} \\
		&=& \sum_{C\in\mathcal{C}; C\textrm{~{\small light}}} \mu(C) \\
		&=& \Pr_{C\sim \mu}\left[\textrm{$C$ is light}\right] \\
		&= & 1- \Pr_{C\sim \mu}\left[\reld(C) > \frac{n^2}{9}\right] \\
		&\ge & 1- \frac{\mathbb{E}_{C\sim \mu}\left[\reld(C)\right]}{n^2/9}
		\textrm{~~~~(by Markov's inequality)}
\end{IEEEeqnarray*}

	So it suffices to show that if a clause $C$ is sampled from $\mathcal{C}$ according to distribution $\mu$, its expected  $\reld(C)$ is small.
	\begin{claim}
		\[ \mathbb{E}_{C\sim \mu}\left[\reld(C)\right] \le \frac{n^2}{10}. \]
	\end{claim}
	\begin{proof}(of claim)
		\begin{align*}
		\mathbb{E}_{C\sim \mu}\left[\reld(C)\right]  & = 
		\sum_{C\in \mathcal{C}: \mu(C)\neq 0} \mu(C) \frac{|C\cap X_5|}{|C\cap X_3|} \\
		& =     \sum_{C\in \mathcal{C}: \mu(C)\neq 0} \frac{|C\cap X_5|}{2|X_3|}
		& \textrm{~~~(each row in $X_3 \times \mathcal{C}$ has exactly 2 1s)} \\
		& = \frac{1}{2|X_3|} \sum_{C\in \mathcal{C}: \mu(C)\neq 0} |C\cap X_5| \\
		& \le \frac{4|X_5|}{2|X_3|}
		& \textrm{~~~(each row in $X_5 \times \mathcal{C}$ has exactly 4 1s)} \\
		& =  \frac{2{n\choose 5}}{{n \choose 3}}
		& \textrm{~~~(by proposition~\ref{prop:Xi-counts})} \\
		& \le   \frac{n^2}{10}.
		\end{align*}
	\end{proof}
	With this claim established, the proof of the Lemma is complete.
\end{proof}

Now we need to show that light clauses cannot contribute many 1s,
Lemma~\ref{lem:light-contribution-low}. We will first obtain, for any
$C\in \mathcal{C}$, estimates for $|C\cap X_3|$ and $|C\cap X_5|$ in
terms of the width $w(C)$ of $C$; Lemma~\ref{lem:N3-N5-estimates}
below. Then we will show that if $C$ is light, then it is wide;
Lemma~\ref{lem:light-wide}. Putting these together will prove
Lemma~\ref{lem:light-contribution-low}.

To state Lemmas~\ref{lem:N3-N5-estimates},\ref{lem:light-wide} we
first need to discuss a suitable subgraph of $G$.  Consider a clause
$C\in\mathcal{C}$ with non-empty $C\cap X_3$. Since
$\viol_{\mathcal{C}} = \viol_F-1$, no assignment in $X_1$ falsifies
$C$.  We rewrite the system $IX= b$ as $I' X' + I_C X_C = b$, where
$X_C$ are the variables fixed in cube $C$ (to $a_C$, say). So $I'X' =
b + I_Ca_C$.  An assignment $a$ is in $C\cap X_r$ iff it is of the
form $a'a_C$, and $a'$ falsifies exactly $r$ equations in $I'X' = b'$
where $b'= b+ I_Ca_C$. This is a system for the subgraph $G_C$ where
the edges in $X_C$ have been deleted. This subgraph may not be
connected, so we cannot use our size expressions from
Proposition~\ref{prop:Xi-counts} directly. Consider the vertex sets
$V_1, V_2, \ldots $ of the components of $G_C$. The system $I'X'=b'$
can be broken up into independent systems; $I'(i)X'(i) = b'(i)$ for
the $i$th connected component. Say a component is odd-charged if $\sum_{j\in
	V_i} b'(i)_j \equiv 1 \bmod 2$, even-charged otherwise. Let $|V_i|=n_i$ and
$|E_i|=m_i$. Any $a'$ falsifies an odd/even number of equations in an
odd-charged/even-charged component.

Pick any $a'\in C\cap X_3$; at least one such assignment exists by
assumption.  It must falsify three equations overall, so $G_C$ must
have either one or three odd-charged components. If it has only one
odd-charged component, then there is another assignment in $C$
falsifying just one equation (from this odd-charged component), so
$C\cap X_1 \neq \emptyset$, a contradiction. Hence $G_C$ has exactly
three odd-charged components, with vertex sets $V_1, V_2, V_3$ of
sizes $n_1,n_2,n_3$ respectively, and overall $k \ge 3$ components.

We now estimate $|C\cap X_3|$ and $|C\cap X_5|$ in terms of these
parameters $n_1,n_2,n_3,k,w(C)$, where $w(C)$ denotes the width of the
clause $C$. Recall that $m=nd/2$ is the number of edges in $G$ and
hence the number of variables in $F$.

\begin{lemma}\label{lem:N3-N5-estimates}
	If  a clause $C\in \mathcal{C}$ has   $|C\cap X_3| \neq 0$, then 
	$|C\cap X_3| =
	n_1n_2n_32^{m-w(C)-n+k}$ and
	\[ |C\cap X_5| \ge n_1n_2n_32^{m-w(C)-n+k} \left( \frac{1}{3}\sum_{i=1}^k{n_i -1 \choose 2}    \right).\]
\end{lemma}
\begin{proof}
	An $a \in C \cap X_3$ falsifies exactly one equation in
	the subsystems $I(1), I(2), I(3)$ corresponding to the odd-charged components of $G_C$. We thus arrive at the expression
	\[|C\cap X_3| = \Biggl(\prod_{i=1}^3n_i 2^{m_i-n_i+1} \Biggr)
	\Biggl(\prod_{i\ge 4} 2^{m_i-n_i+1} \Biggr)
	= n_1n_2n_3 2^{m-w(C) - n + k}.
	\]
	Similarly, an $a\in C\cap X_5$ must falsify five equations overall. One each
	must be from $V_1, V_2, V_3$. The remaining 2 must be from the same
	component. Hence
	\begin{align*}
	|C\cap X_5| &=
	\left({n_1\choose 3}n_2n_3 
	+
	n_1{n_2\choose 3}n_3 
	+
	n_1n_2{n_3\choose 3}\right) 2^{m-w(C) - n + k} 
	\\
	&+
	n_1n_2n_3\sum_{i=4}^k {n_i \choose 2} 2^{m-w(C) - n + k} 
	\\
	&\ge n_1n_2n_3 2^{m-w(C) - n + k}
	\left( \frac{1}{3}\sum_{i=1}^k {n_i-1 \choose 2} \right) 
	\end{align*}
\end{proof}

Now we use the structure and parameters of $G_C$ to show that light
clauses must be wide.
\begin{lemma}\label{lem:light-wide}
	For any clause $C\in \mathcal{C}$, if
	$\reld(C) = \frac{|C\cap X_5|}{|C\cap X_3|} \le \frac{n^2}{9}$,
	then $w(C) \ge \frac{cn}{10}$.
\end{lemma}
\begin{proof}
	Each literal in $C$
	removes one edge from $G$ while constructing $G_C$.  Counting the
	sizes of the cuts that isolate components of $G_C$, we count each
	deleted edge twice.  So
	\[
	2w(C) = \sum_{i=1}^k |\delta(V_i,V\setminus V_i)| 
	= \sum_{i: n_i \le n/2} \underbrace{|\delta(V_i,V\setminus V_i)|}_{Q1} +
	\sum_{i: n_i > n/2} \underbrace{|\delta(V_i,V\setminus V_i)|}_{Q2} 
	\]
	By the $c$-expansion property of $G$, $Q1 \ge cn_i$.  \\
	If $n_i > n/2$, it still cannot be too large because $C$ is light.
	Recall
	\[ \frac{n^2}{9} \ge \frac{|C\cap X_5|}{|C\cap X_3|} \ge  \frac{1}{3}\sum_{i=1}^k {n_i-1 \choose 2} \]
	If any $n_i$ is very large, say larger than $5n/6$, then the
	contribution from that component alone, $\frac{1}{3}{n_i-1 \choose
		2}$, will exceed $\frac{n^2}{9}$. So each $n_i\le
	5n/6$. Thus even when $n_i > n/2$, we can conclude that $n_i/5 \le n/6
	\le n-n_i < n/2$. By expansion of $V \setminus V_i$, we have $Q2 \ge
	c(n-n_i) \ge cn_i/5$.
	\begin{eqnarray*}
		2w(C) 
		&=& \sum_{i: n_i \le n/2} \underbrace{|\delta(V_i,V\setminus V_i)|}_{Q1} +
		\sum_{i: n_i > n/2} \underbrace{|\delta(V_i,V\setminus V_i)|}_{Q2}  \\
		&\ge & \sum_{i: n_i \le n/2} cn_i + \sum_{i: n_i > n/2} \frac{c n_i}{5}
		\ge cn/5
	\end{eqnarray*}
	Hence $w(C) \ge cn/10$ as claimed. 
\end{proof}

Now we have all that is needed to prove
Lemma~\ref{lem:light-contribution-low}.
\begin{proof}(of Lemma~\ref{lem:light-contribution-low})
	Let $C$ be a light clause. As discussed above, let $G_C$ be the
	subgraph of $G$ where edges whose variables are set by $C$ are
	deleted, let $k$ be the number of components of $G_C$, and let
	$n_1,n_2,n_3$ be the number of vertices in the three odd-charged
	components.
	\begin{align*}
	|C\cap X_3| & = n_1n_2n_32^{m-w(C)-n+k}
	& \textrm{~~~(by Lemma~\ref{lem:N3-N5-estimates})} \\
	& = \frac{n_1n_2n_32^{m-w(C)-n+k}}{{n\choose 3}2^{m-n+1}} \times |X_3|
	& \textrm{~~~(by Proposition~\ref{prop:Xi-counts})} \\
	& = \frac{n_1n_2n_3}{{n\choose 3}}2^{k-w(C)-1} \times |X_3| \\
	& \le 6 \times 2^{n-w(C)-1} \times |X_3| = 3 \cdot  2^{n-w(C)} \cdot |X_3|\\
	& \le 3 \cdot  2^{n-cn/10} \cdot  |X_3|
	& \textrm{~~~(by Lemma~\ref{lem:light-wide})} \\
	& = \frac{3|X_3|}{2^{n(0.1c-1)}}
	& \textrm{~~~as claimed.} \\
	\end{align*}
This completes the proof of Theorem~\ref{thm:ssc-Tseitin}.
\end{proof}

\paragraph{Remark}
As noted in Section~\ref{sec:defs}, the
\SSC\ proof system can be viewed algebraically as a subsystem of
Sherali-Adams, for which this lower bound is already known.  However,
our proof is specific to the \SSC\ proof system, where all the
multipliers for the axiom polynomials are $-1$. This is implicit in
our proof; we use the equation $\viol_{\mathcal{C}} = \viol_F-1$, and
thus we assume that the axiom polynomials from $F$ are multiplied only
by $-1$.

\subsection{Lifting degree lower  bounds to size}
\label{sec:ssc-lifting}

We describe a general technique to lift lower bounds on width, or conical junta
degree, to  lower bounds on combinatorial size for \SSC. 
This is an adaptation of the well-known xorification technique of
Alekhnovich and Razborov (see~\cite{Ben-Sasson09SizeSpaceTradeoffs}),
which also consists of applying a random restriction to a formula
composed with parity.

\begin{theorem}
  \label{thm:ssc-lifting}
  Let $d$ be the minimum width, or conical junta degree, of a \SSC{} refutation of an
  unsatisfiable CNF formula $F$. Then every \SSC{} refutation of
  $F\circ \oplus$ has combinatorial size $ \exp(\bigomega{d})$.
\end{theorem}

Before proving this theorem, we establish two lemmas.  For a function
$h\colon\allowbreak\{0,1\}^n \to \Re$, define the function
$h\circ\oplus\colon\{0,1\}^{2n}\rightarrow \Re$ as
$(h\circ\oplus)(\alpha_1,\alpha_2)=h(\alpha_1\oplus\alpha_2)$, where
$\alpha_1,\alpha_2\in\{0,1\}^n$ and the $\oplus$ in
$\alpha_1\oplus\alpha_2$ is taken bitwise.

\begin{lemma}
  \label{lem:viol-commutes}
  $\viol_F(\alpha_1 \oplus \alpha_2) = \viol_{F \circ \oplus
  }(\alpha_1,\alpha_2)$.
\end{lemma}
\begin{proof}
  Fix assignments $\alpha_1$, $\alpha_2$ and let
  $\alpha = \alpha_1 \oplus \alpha_2$. We claim that for each clause
  $C \in F$ falsified by $\alpha$ there is exactly one clause
  $D \in F \circ \oplus$ that is falsified by
  $\alpha_1\alpha_2$. Indeed, by the definition of composed formula
  the assignment $\alpha_1\alpha_2$ falsifies $C \circ \oplus$, hence
  the assignment falsifies some clause $D \in C \circ
  \oplus$. However, the clauses in the CNF expansion of
  $C \circ \oplus$ have disjoint subcubes, hence $\alpha_1\alpha_2$
  falsifies at most one clause from the same block. Observing that if $\alpha$ does not falsify $C$, then $\alpha_1\alpha_2$ does not falsify any clause in $C \circ
  \oplus$ completes the proof.
\end{proof}

Note that Lemma~\ref{lem:viol-commutes} may not be true for gadgets other than $\oplus$.

\begin{corollary}
\label{cor:viol-commutes}
$\viol_{F\circ\oplus} -1 = ((\viol_F)\circ\oplus)-1 = (\viol_{F} -1)\circ\oplus$.
\end{corollary}
\begin{proof}
  $((\viol_{F} -1)\circ\oplus)(\alpha_1,\alpha_2) =
  (\viol_{F} -1)(\alpha_1\oplus\alpha_2) =
  (\viol_{F})(\alpha_1\oplus\alpha_2) -1 =
  (\viol_{F\circ\oplus})(\alpha_1,\alpha_2) -1 $.
  \end{proof}
\begin{lemma}
  \label{lem:size-degree}
  If $f \circ \oplus$ has a (integral) conical junta of size $s$, then $f$ has
  a (integral) conical junta of degree $d=\bigoh{\log s}$. 
\end{lemma}
\begin{proof}
  Let $J$ be a conical junta of size $s$ that computes
  $f \circ \oplus$. Let $\rho$ be the following random restriction:
  for each original variable $x$ of $f$, pick $i\in\set{0,1}$ and
  $b\in\set{0,1}$ uniformly and set $x_i=b$. Consider a term $C$ of
  $J$ of degree at least $d>\log_{4/3} s$. The probability that $C$ is
  not zeroed out by $\rho$ is at most $(3/4)^{d}<1/s$, hence by a
  union bound the probability that the junta $\restrict{J}{\rho}$ has
  degree larger than $d$ is at most $s \cdot (3/4)^{d} <1$. Hence
  there is a restriction $\rho$ such that $\restrict{J}{\rho}$ is a
  junta of degree at most $d$, although not one that computes $f$. Since
  for each original variable $x$, $\rho$ sets exactly one of the
  variables $x_0,x_1$, flipping the appropriate surviving
  variables---those where $x_i$ is set to 1---gives a junta of degree
  at most $d$ for $f$.  \end{proof}

Now we can prove Theorem~\ref{thm:ssc-lifting}.
\begin{proof}
We prove the contrapositive: if $F \circ \oplus$ has a \SSC\ proof of
combinatorial size $s$, then there is an integral conical junta for $g = \viol_F -1$ of degree
$\bigoh{\log s}$.

Let $H$ be the collection of cubes in the \SSC\ proof for
$F\circ\oplus$.  So $\viol_{F\circ\oplus} -1 = \viol_H$. By
Corollary~\ref{cor:viol-commutes}, there is an integral conical junta for
$(\viol_{F}-1)\circ \oplus$ of size $s$. By
Lemma~\ref{lem:size-degree} there is an integral conical junta for $\viol_{F}-1$
of degree $\bigoh{\log s}$.  \end{proof}

\paragraph{Recovering the Tseitin lower bound:}
This theorem, along with the $\bigomega{n}$ conical junta degree lower
bound of~\cite{GJW18ExtensionComplexity}, yields an exponential lower
bound for the \SSC\ and \maxresw\ refutation size for Tseitin
contradictions. %
However, this construction duplicates every edge of the original graph and therefore does not give a lower bound for all expanders.

\paragraph{A candidate for separating \res\ from \SSC:}
We conjecture that the \SSC\ degree of the pebbling contradiction on
the pyramid graph, or on a minor modification of it (a stack of butterfly networks, say, at the base of a pyramid), is
$n^{\bigomega{1}}$. This, along with Theorem~\ref{thm:ssc-lifting}
would imply that $F\circ\oplus$ is hard for \SSC, thereby separating
it from \res. However we have not yet been able to prove the desired
degree lower bound. We do know that \SSC{} degree is not exactly the same
as  \res{} width -- for small examples, a brute-force computation has shown \SSC{} degree to be strictly larger than \res{} width. 

\section{Discussion}
\label{sec:conclusion}

We placed \maxres\ and \maxresw\
in a propositional proof complexity frame and
compared it to more standard proof systems, showing that MaxResW is
between tree-like resolution (strictly) and resolution. With the goal
of also separating MaxRes and resolution we devised a new lower bound
technique, captured by \SSC, and proved lower bounds for MaxRes
without relying on \res\ lower bounds.

Perhaps the most conspicuous problem left open in this paper is whether our conjecture
that pebbling contradictions composed with XOR separate \res{} and
\SSC{} holds. (Very recently, in \cite{FGGR-ITCS22}, this has been resolved by showing precisely such a separation.)
It remains open to show that MaxRes simulates
\tres{} -- or even MaxResW -- or that they are incomparable
instead.

\subsection*{Acknowledgments}
Part of this work was done when the last author was at TIFR, Mumbai, India. 
Some of this work was done in the conducive academic environs of
the Chennai Mathematical Institute (during the CAALM workshop of
CNRS UMI ReLaX, 2019), Banff
International Research Station BIRS (seminar 20w5144) and Schloss
Dagstuhl Leibniz Centre for Informatics (seminar 20061).  The authors  thank Susanna de Rezende, Tuomas Hakoniemi, and Aaron Potechin  for useful discussions.

\bibliography{GSbib}
\bibliographystyle{ACM-Reference-Format}

\end{document}